**Hemolithin: a Meteoritic Protein containing Iron and Lithium.**

Malcolm. W. McGeoch[1], Sergei Dikler[2]
and Julie E. M. McGeoch[3*].

[1] PLEX Corporation, 275 Martine St., Suite 100, Fall River, MA 02723, USA.
[2] Bruker Scientific LLC, 40 Manning Rd, Billerica MA 01821.
[3] Department of Molecular and Cellular Biology, Harvard University, 52 Oxford St., Cambridge MA 02138, USA.
*Corresponding author. E-mail: mcgeoch@fas.harvard.edu

## ABSTRACT

This paper characterizes the first protein to be discovered in a meteorite. Amino acid polymers previously observed in Acfer 086 and Allende meteorites [1,2] have been further characterized in Acfer 086 via high precision MALDI mass spectrometry to reveal a principal unified structure of molecular weight 2320 Daltons that involves chains of glycine and hydroxy-glycine residues terminated by iron atoms, with additional oxygen and lithium atoms. Signal-to-noise ratios up to 135 have allowed the quantification of iron and lithium in the various MALDI fragments via the isotope satellites due to their respective minority isotopic masses $^{54}Fe$ and $^{6}Li$. Analysis of the complete spectrum of isotopes associated with each molecular fragment shows $^{2}H$ enhancements above terrestrial averaging 25,700 parts per thousand (sigma = 3,500, n=15), confirming extra-terrestrial origin and hence the existence of this molecule within the asteroid parent body of the CV3 meteorite class. The molecule is tipped by an iron-oxygen-iron grouping that in other terrestrial contexts has been proposed to be capable of absorbing photons and splitting water into hydroxyl and hydrogen moieties.

## INTRODUCTION

Although individual amino acids have been found in abundance in carbonaceous meteorites there have been only two reports of polymers of amino acids, the first being of di-glycine [3], and the more recent report [1] being of large polymers of mainly glycine in the CV3 class carbonaceous chondrite Allende. In follow-on work to [1] a 4641Da molecular entity was discovered in Allende and Acfer 086 [2] with extra-terrestrial isotope enhancement that confirmed that these unexpected molecules were not artifacts due to terrestrial contamination. In the present study a state-of-the-art mass spectrometer [4] generated much-improved signal-to-noise ratios that allowed us to discover via the $^{54}Fe$ isotope satellite that iron was also present and bonded to the glycine and hydroxy-glycine, in specific arrangements. Also, via its $^{6}Li$ satellite, lithium was found to be a standard component. Much of the present paper is devoted to the analysis via isotope satellites and the reconstruction of a few core molecular entities that give rise to all of the observed peaks in a mass spectrum. When the prior 4641Da observations are also taken into account, the most probable core molecule has a mass of 2320Da, containing two glycine strands each of length 16 residues, but variants with 15 and 17 residues also appear to be present. Following the presentation of methods, the smaller molecules with m/z < 1000, generated in intense laser conditions, are reviewed. Less severe conditions are then applied in order to go to masses as high as 2364 so as to understand the polymer structure. The discussion aims to place this molecule in context, whether its origin was in the proto-solar disc, or in molecular clouds well before the solar system condensed.

## SIGNIFICANCE

This is the first report of a protein from any extra-terrestrial source. Room temperature extracts from micron-sized meteorite particles contain polymers of amino acids with a definite chain length centered at 16 residues. Analysis via iron and lithium isotope satellites in mass spectrometry reveals a novel protein motif with iron atoms closing out the ends of anti-parallel peptide chains composed of glycine. Very high $^{2}H$ content indicates proto-solar disc or molecular cloud origin. $FeO_3Fe$ groups at each end are of a type that could split $H_2O$ upon absorption of photons. The existence of a unique chain length suggests that there could be a functionality conferring a replication advantage.

## METHODS

### Sample preparation

In summary: Micron particles of Acfer-086 were Folch extracted [1,5] into 4 phases that we label P1 –> P4 : 1) top polar phase (P1); 2) above interphase (P2); 3) interphase (P3); 4) bottom chloroform phase (P4). Specifically, Acfer-086, a CV3 meteorite sample from Harvard Mineralogical and Geological Museum (Source: Agemour, Algeria, found 1989-90, TKW 173g) was delivered from the Museum in a sealed container to the Hoffman, Earth and Planetary Science clean room.

**Meteorite Etch**: In a clean room extractor hood, at room temperature with high airflow, the samples were hand held with powder-free nitrile rubber gloves while being etched to a total depth of 6mm with diamond burrs. The diamonds had been vacuum-brazed at high temperature onto the stainless-steel burr shafts to avoid the presence of glue of animal origin and organics in general. Etching on a fracture face (not an original exterior weathered face) was via slow steady rotation of a burr under light applied force via a miniature stepper motor that did not have motor brushes and did not contribute metal or lubricant contamination to the clean room. Two shapes of burr were used, the larger diameter type, in two stages, to create a pit of diameter 6mm and depth 6mm, and the smaller conical burr to etch a finely powdered sample, of approximately 1μm particles, from the bottom of the pit without contacting the sides. After each stage the pit contents were decanted by inversion and tapping the reverse side and a new burr was used that had been cleaned by ultrasonics in deionized distilled (DI) water followed by rinsing in DI water and air-drying in a clean room hood. The powder from the third etch was decanted with inversion and tapping into a glass vial. Sample weights were in the range 2-8mg.

### Folch Extraction of polymer amide from meteorites

Solvents chloroform/methanol/water were added to the glass vials containing the etched particles in the ratio 70/20/10 per the Folch extraction protocol [5] to give a meteorite particle/Folch extract solvent concentration of 2-3mg/ml. Chloroform is pipetted first into the glass bottle containing the meteorite particles, followed by methanol and then water to get the phases established. Gentle swirling of the contents results in the meteorite particles locating at the interface between the bottom phase (mainly chloroform) and top phases (methanol and water). As the particles solvate, a torus topology forms at the interphase of the chloroform and methanol/water phases [1]. After 1 day of extraction at room temperature the Folch extract phases were pipetted off as 50-100μl aliquots from the top downwards through the phases, each extract being labeled P1 through P4 in the above order of phases. In a second identical extraction experiment, only the above-interphase P2 was collected in view of its larger complement of the molecule in question. The residual phases P3 and P4 from this second experiment were re-established by addition of chloroform, methanol and water into a third experiment that was kept at room temperature for 7 more days (totaling 8). A flocculent precipitate appeared at the interphase and was pipetted off to become sample P2a. These aliquots were then analyzed by MALDI mass spectrometry.

**Sample Preparation for MALDI MS.**

The extracts were prepared with three MALDI matrices: α-cyano-4-hydroxycinnamic acid (CHCA), sinapinic acid (SA) and 2,5-dihydroxybenzoic acid (DHB). The CHCA solution was prepared at 10 mg/mL in 50% acetonitrile, 0.1% TFA. The saturated SA solution was prepared in ethanol. The second SA solution for double layer preparation at 20 mg/mL was prepared in 50% acetonitrile, 0.1% TFA. The DHB solution at 45 mg/mL was prepared in 30% acetonitrile, 0.1% TFA. The extract aliquots were mixed 1:1 with the CHCA solution or the DHB solution and 1 μL of the mixture was spotted in duplicate on a new, never used MTP 384 Ground Steel MALDI plate. The double layer SA preparation was accomplished by spotting 0.5 μL of the saturated SA solution first followed by 1 μL of 1:1 mixture of extract and the second SA solution at 20 mg/mL on the same MALDI plate.

**MALDI-TOF and MALDI-TOF/TOF measurements**

All spectra were acquired on a Bruker rapifleX MALDI-TOF/TOF system [4] equipped with a scanning Smartbeam 3D laser operating at 10 kHz. The instrument was operated in reflector positive ion mode. The 10-bit digitizer was set to a sampling rate of 5.00 GS/s. The ion source voltage 1 (acceleration voltage) was set to 20.00 kV, the ion source voltage 2 (extraction voltage) was at 17.44 kV and lens voltage was at 11.60 kV. The pulsed ion extraction time was set to 160 ns. The reflector voltage 1 was at 20.82 kV, reflector voltage 2 at 1.085 kV and reflector voltage 3 at 8.60 kV. Four thousand laser shots were acquired per spectrum.

The MS/MS spectra were acquired in TOF/TOF mode and 20000 laser shots were added per spectrum.

## RESULTS

**Small molecular fragments: finding the dominant constituents.**

With the MALDI parameters optimized for fragmentation in the 1-1000 m/z range, a study was performed on phase P2, which already had been established as the phase carrying most of the molecules under study. The 1-1000 m/z range is cluttered with matrix clusters that are difficult to distinguish from real signals, although most matrix clusters can be predicted [6,7]. Our samples had a relatively high content of alkali atoms (described in [2]), which made clustering very severe, but rather than risk contamination by processing to remove alkalis we adopted a different stratagem. Instead of subtracting from a spectrum all the calculated clusters as in our prior analysis [1], we ran the same sample separately with each of the matrices CHCA, SA and DHB, and only selected the peaks that appeared in all three spectra for analysis, guaranteeing the elimination from this list of essentially all clusters. All such peaks for sample P2 in the 1-1000 range, with peak counts more than 15 times greater than noise, are listed in Table 1, together with their proposed composition in terms of only glycine and hydroxy-glycine amino acid residues, which have mono-isotopic residue masses 57.02146 and 73.01638 respectively. The matrix that yielded the largest peak count out of the three is listed. A large proportion of the peaks (16 out of 36) could be constructed from small polymers containing only these two residues provided that we included the possibility of three types of termination:

a) the bare combination of residues;

b) the combination of residues plus one proton;

c) the combination of residues plus an "aqueous" termination 'OH' + 'H'.

Other terminations that were close to a match are denoted by italics in the table, but these molecular species were not counted. By enumeration of the possibilities we determined that the probability of achieving a compositional match with any of these (type a, b or c) terminations by chance was 0.256, and found using the binomial distribution that the chance of our having obtained 16 or more matches with exactly the terminations (type a, b or c) out of 36 peaks was only 0.010. This confirmed the presence of these two amino acids, but did not exclude other possibilities. The average fragment length in this set was 4.5 residues. Lower temperature conditions within the MALDI plasma would allow larger molecular fragments to survive, as found in the following section.

**Table 1. Low molecular weight fragments from P2 with assignments**

| Observed m/z | Peak counts | Matrix | Assignment (Gly, $Gly_{OH}$) + t | $^{2}H$ fitted | $^{15}N$ preset | Calc. m/z |
|---|---|---|---|---|---|---|
| 57.074 | 132,670 | CHCA | (1,0) | | | 57.021 |
| 73.002 | 6,669 | DHB | (0,1) | | | 73.016 |
| 74.102 | 678,662 | CHCA | (0,1) + H | | | 74.024 |
| 114.096 | 101,559 | CHCA | (2,0) | | | 114.043 |
| 128.113 | 734,815 | CHCA | - | | | - |
| 130.166 | 172,225 | CHCA | (1,1) | | | 130.038 |
| 161.046 | 746,629 | DHB | - | | | - |
| 169.082 | 1,238,632 | CHCA | - | | | - |
| 172.045 | 1,511,547 | CHCA | (3,0) + H | | | 172.072 |
| 185.078 | 239,372 | CHCA | - | | | - |
| 197.071 | 299,125 | CHCA | - | | | - |
| 237.092 | 313,018 | SA | (0,3) + 'OH' + H | | | 237.060 |
| 244.271 | 260,294 | CHCA | (3,1) | | | 244.081 |
| 256.270 | 63,527 | CHCA | - | | | - |
| 272.300 | 72,300 | CHCA | - | | | - |
| 279.102 | 555,493 | CHCA | (2,2) + 'OH' + *2H* | | | 279.094 |
| 286.280 | 159,152 | CHCA | (5,0) + H | | | 286.115 |
| 293.134 | 248,245 | CHCA | (0,4) + H | | | 293.073 |
| 295.099 | 89,961 | SA | - | | | - |
| 304.312 | 454,322 | SA | (5,0) + 'OH' + *2H* | | | 304.126 |
| 316.329 | 178,473 | DHB | (3,2) - *H* | 30,000 | 1,015 | 316.089 |
| 332.342 | 239,731 | SA | (2,3) - *H* | 30,000 | 1,015 | 332.084 |
| 337.129 | 219,817 | CHCA | - | | | - |
| 375.142 | 43,128 | SA | (4,2) + H | | | 375.126 |
| 401.086 | 504,156 | CHCA | (7,0) + *2H* | 30,000 | 1,015 | 401.166 |
| 433.172 | 147,707 | SA | (6,1) + 'OH' + H | | | 433.156 |
| 449.168 | 316,351 | SA | (5,2) + 'OH' + H | | | 449.151 |
| 477.147 | 27,497 | SA | (8,0) + 'OH' + *4H* | | | 477.206 |
| 487.315 | 145,270 | DHB | (6,2) - *H* | 30,000 | 1,015 | 487.154 |
| 493.321 | 274,754 | CHCA | (7,1) + 'OH' + *4H* | | | 493.201 |
| 503.091 | 53,862 | SA | - | | | - |
| 509.296 | 213,767 | CHCA | (6,2) + 'OH' + *4H* | | | 509.196 |
| 515.302 | 105,500 | CHCA | (9,0) + *2H* | | | 515.209 |
| 531.275 | 77,837 | CHCA | (9,0) + 'OH' + H | | | 531.204 |
| 547.247 | 38,571 | CHCA | (8,1) + 'OH' + H | | | 547.199 |
| 563.217 | 25,057 | CHCA | (7,2) + 'OH' + H | | | 563.194 |

**Spectra in the high m/z range 1 - 5,000**

Using less intense laser conditions and two matrices CHCA and SA, mass spectra were collected in the m/z range 1-5000 on phases P2 and P3. These four data sets were cross-correlated and only the peaks that appeared in all four (20 in number) were included in a first list of candidate species that is reproduced in Table 2. This selection approach ensured that the peaks were not matrix artifacts, i.e. polymers of matrix alone or complexes of matrix with the molecule under study, because the different matrix masses (189 for CHCA and 224 for SA) would give rise to different complexes, and these would not cross-correlate between the spectra. For continuity with our previous analyses [1,2] we show in Table 2 the amino acid composition of glycine, hydroxy-glycine and alanine that we would have assigned if this had been the total of information available. The average residue count implied by this assignment was 23, a large increase above the previous average of 11 residues that we had observed [2].

**Table 2. List of peaks that appeared in all four tests, with the (now outdated) assignment in terms of three amino acids and an aqueous termination.**

| Integer name of peak | Average m/z (four traces) | Historical assignment according to [1] in terms of Gly, $Gly_{OH}$, Ala, and a termination. | | | | Calc. mass based on assignment |
|---|---|---|---|---|---|---|
| | | Gly | $Gly_{OH}$ | Ala | termination | |
| 693 | 693.256 | 8 | 3 | 0 | 18 | 693.231 |
| 947 | 947.309 | 15 | 1 | 0 | 18+H | 947.357 |
| 963 | 963.274 | 14 | 2 | 0 | 18+H | 963.352 |
| 1001 | 1001.502 | 16 | 0 | 1 | 18 | 1001.391 |
| 1011 | 1011.567 | 11 | 5 | 0 | 18+H | 1011.336 |
| 1027 | 1027.542 | 10 | 6 | 0 | 18+H | 1027.331 |
| 1033 | 1033.537 | 14 | 2 | 1 | 18 | 1033.381 |
| 1417 | 1417.513 | 22 | 1 | 1 | 18+H | 1417.544 |
| 1433 | 1433.486 | 21 | 2 | 1 | 18+H | 1433.539 |
| 1449 | 1449.532 | 20 | 3 | 1 | 18+H | 1449.534 |
| 1483 | 1483.864 | 18 | 6 | 0 | 18+H | 1483.503 |
| 1503 | 1503.741 | 21 | 2 | 2 | 18 | 1503.568 |
| 1519 | 1519.697 | 20 | 3 | 2 | 18 | 1519.563 |
| 1535 | 1535.683 | 19 | 4 | 2 | 18 | 1535.558 |
| 1551 | 1551.650 | 18 | 5 | 2 | 18 | 1551.553 |
| 1567 | 1567.729 | 17 | 6 | 2 | 18 | 1567.548 |
| 1639 | 1639.630 | 17 | 6 | 3 | 18+H | 1639.593 |
| 1995 | 1995.993 | 27 | 6 | 0 | 18 | 1995.688 |
| 2012 | 2013.019 | 26 | 7 | 0 | 18+H | 2012.691 |
| 2124 | 2124.882 | 28 | 6 | 1 | 18+H | 2124.755 |
| | sum | 362 | 76 | 19 | | |

It was remarkable that all 20 peaks could be immediately assigned in this manner, which reinforced the certainty that the prior molecular type was again being observed, this time at higher signal-to-noise ratio. The assignments of Table 2, however, are now incorrect in light of the data to be presented below.

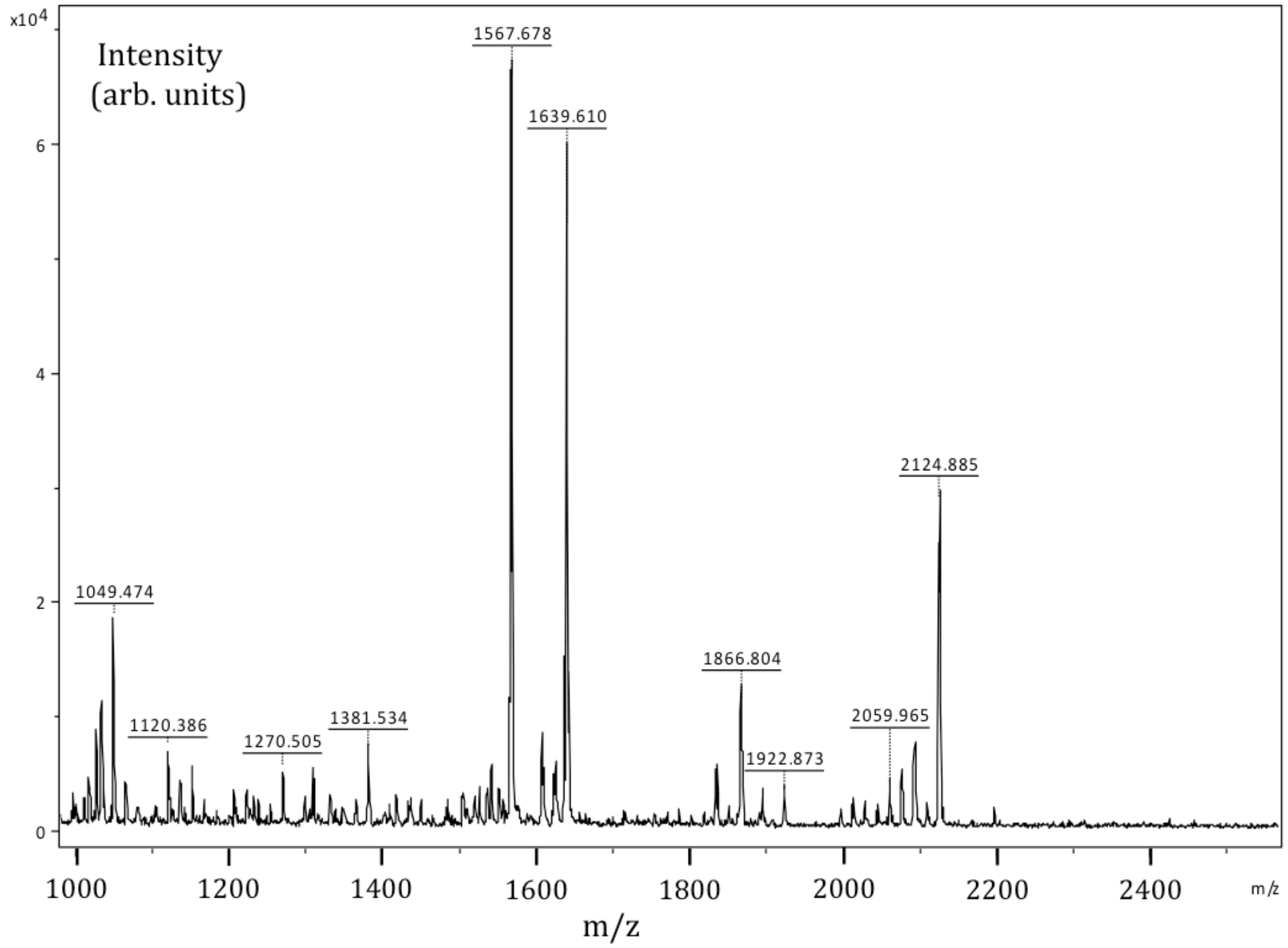

**Figure 1. Major m/z peaks in the P2 mass spectrum with matrix CHCA. The spectrum stops at m/z 2124 and in this range of m/z with matrix CHCA almost all peaks are related to the molecule under study.**

A general mass spectrum from P2 is shown in Figure 1. Using matrix CHCA no major peaks are seen above m/z 2124.885, which we shall refer to by its integer part as just "2124". Below this there are dominant peaks at 1866, 1639, 1567, and 1049. One of the lowest mass peaks with high intensity (not shown in Figure 1) is at m/z 730, shown in detail in Figure 2. It was chosen in order to illustrate the isotope satellites in a simple low molecular weight species with only one iron atom. Iron can be shown to be present in the ion at m/z 730 by its (-2) and (-1) satellites that relate respectively to the (0) and (1) satellites in the abundance ratio 0.05845 : 0.91754 of iron isotopes $^{54}Fe$ and $^{56}Fe$. There are in addition two less abundant stable iron isotopes with satellites on the high mass side, $^{57}Fe$ with abundance 0.0212 and $^{58}Fe$ at 0.0028, but these are not separately visible above the many other heavy isotope contributions at locations (1), (2), etc.

There is a characteristic (-2), (-1), (0), (1) pattern that is seen throughout the present study, with varying satellite amplitudes indicating the presence of from 1 to 4 iron atoms in a species. As an example of this the high intensity '1567' peak at S/N = 135 is shown in Figure 3. The 1567 isotope fit shown in Figure 4, with details in Table 3, shows that it has 3 iron atoms. The calculation of an isotope pattern from a given trial composition is described in S1. The signal-to-noise level around a main peak has to be high enough to distinguish between different numbers of iron atoms. The 54/56 abundance ratio is 0.064 [8], so a single iron atom gives a (-2)/(0) ratio of approximately 6.4% and the signal-to-noise ratio therefore has to exceed about 30 to distinguish between, say, one and two iron atoms, or three and four of them.

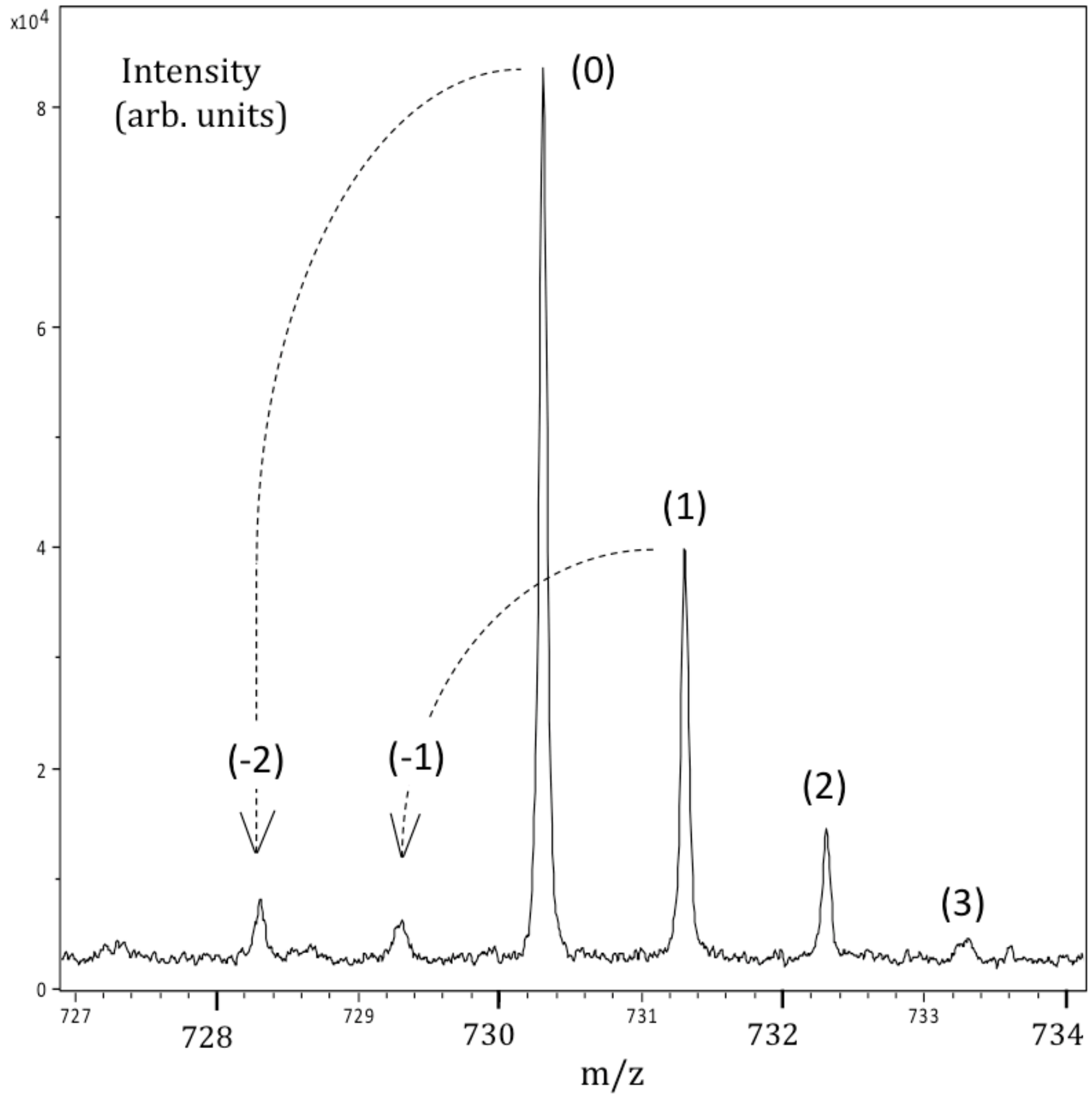

**Figure 2. Detail of the peak at m/z 730.306 in the phase P2 spectrum with CHCA. The labeling of isotope peaks is illustrated, together with the mechanism whereby Fe with majority $^{56}Fe$ in the molecular peaks (0), (1), (2) etc. contributes to satellites 2 units to the left via its minority $^{54}Fe$ isotope. Isotope fitting for 730 is listed in S2.**

**The 1567 fragment**

One of the highest intensity species present in the general spectrum (Figure 1) is at m/z 1567, but this is not a primary molecule because it appears in the MS/MS breakdown spectra (discussed below) as part of each of the 1639 and 2124 species. In Figure 4 the 1567 data has been fitted using the composition {(20Gly, 2$Gly_{OH}$, 3Fe) + 7'O' + H} that emerged from the analysis. Only two parameters were varied in order to match the experimental satellite spectrum: a) the number of Fe atoms, and b) the hydrogen isotope excess above terrestrial (method in S1). With terrestrial $^{2}H$ content the spectrum would have been very different at the (1), (2), (3) etc. levels, as illustrated alongside the fitted spectrum in the dotted trace of Figure 4. The level of $^{2}H$ enhancement above terrestrial to fit the 1567 spectrum was 27,500‰ ± 2,500‰ (where the ‰ symbol represents parts per thousand above terrestrial). The fit was performed with a simultaneous $^{15}N$ preset enhancement of 1,015‰ that had been measured by a different method [9]. The modeled structure was derived from two directions: a) the fragments that 1567 breaks into under MS/MS, and b) analysis of the species that contain 1567 in their own MS/MS products, specifically compounds at m/z 1639 and m/z 2124. The relationship between these species is mapped in S4.

**Table 3. Isotope satellite analysis for peak at m/z 1567 of Figure 3. S/N = 135. Modeled structure = (20Gly, 2$Gly_{OH}$, 3Fe) + 7'O' + H.**

| Satellite | (-3) | (-2) | (-1) | (0) | (1) | (2) | (3) |
|---|---|---|---|---|---|---|---|
| Data | 1,427 | 11,721 | 10,794 | 66,519 | 67,450 | 34,637 | 14,141 |
| Fitted $^{2}H$ 27,500 ‰ Preset $^{15}N$ 1,015 ‰ | .004 | .060 | .060 | .314 | .301 | .168 | .068 |
| Normalized data | .007 | .055 | .051 | .314 | .318 | .164 | .067 |
| Calc. terrestrial | .003 | .058 | .037 | .314 | .200 | .083 | .026 |

A similarly high $^{2}H$ enhancement was observed in all peaks to be analyzed in this way, data summarized in S2. It indicates extra-terrestrial origin for the molecules under study.

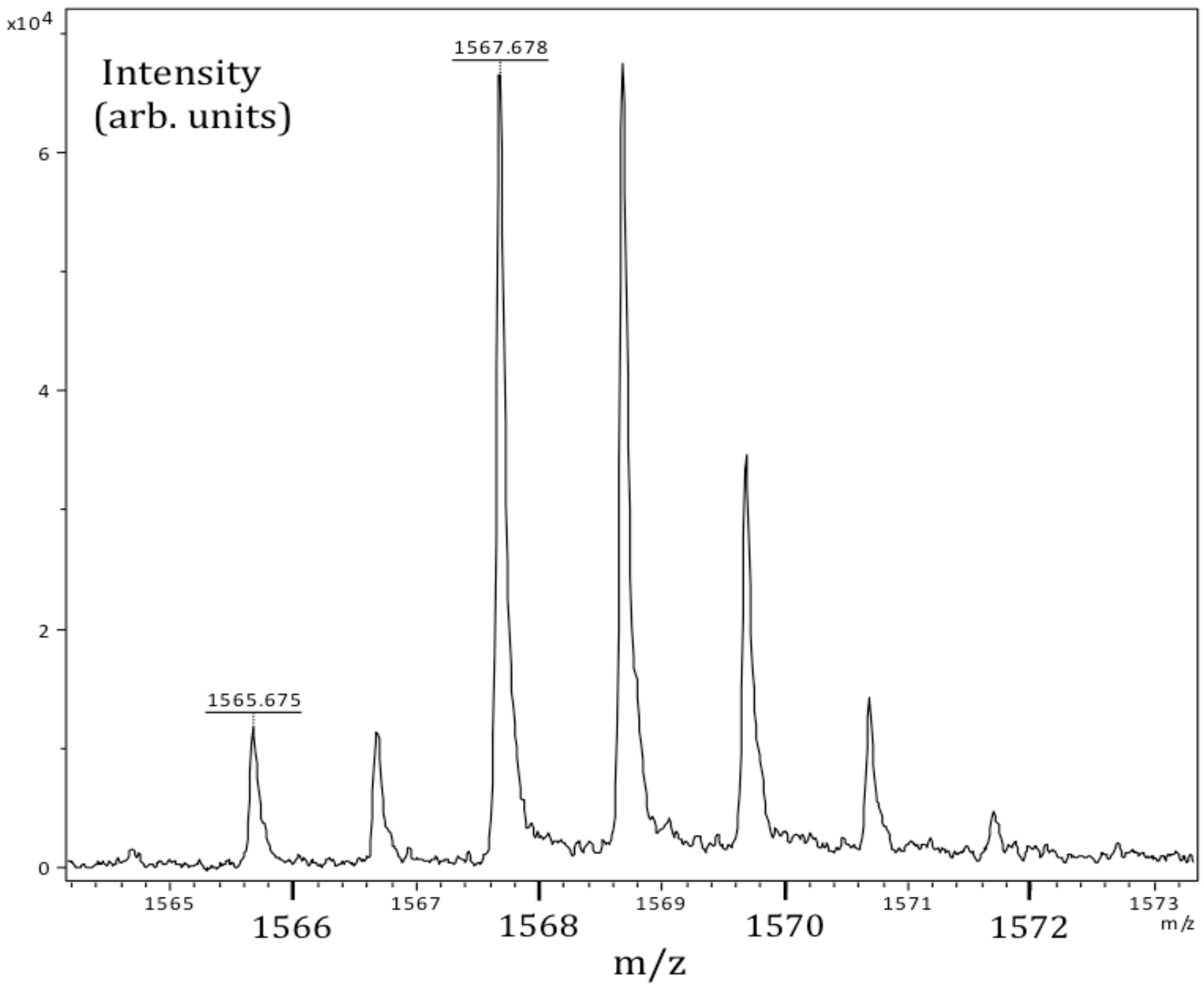

**Figure 3 The 1567.678 m/z peak from phase P2 with matrix CHCA**

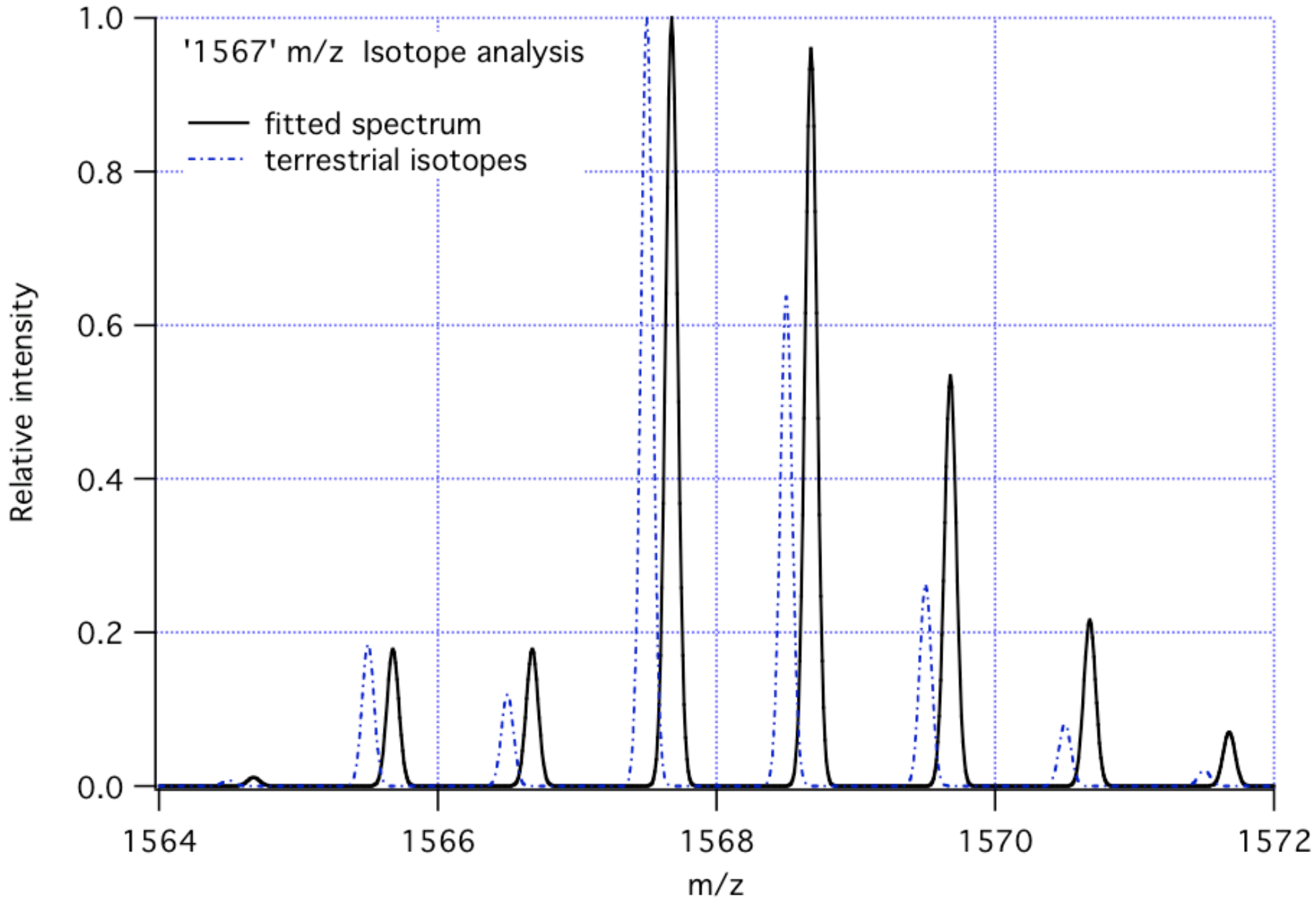

**Figure 4. Fitted isotope spectrum (solid line) to 1567 peak and simulation of the same structure with terrestrial isotope values (dotted line, offset for clarity). Details of the fit are given in Table 3.**

**The 2069 species and others with lithium.**

In addition to the analysis of iron content mainly via the (-2) satellite peak, a measure of lithium content was obtained mainly from the amplitude of the (-1) satellite. This arises from the $^{6}Li$ : $^{7}Li$ abundance ratio of 0.0759 : 0.9241. In a previous analysis there had been evidence for the lithium mass of 7Da in the fitting of polymers to m/z observations [2], but there had not been a strong technique to differentiate between the presence of the pair ($^{16}O$ + $^{7}Li$) and the presence of $^{23}Na$. With higher signal-to-noise ratios the (-1) : (0) satellite ratio allows us to confirm definitely the presence of lithium and to differentiate between one and two lithium atoms at a spectrum signal-to-noise ratio of greater than about 30. The peak at m/z 2069 with S/N = 63 shows the presence of two lithium atoms upon isotope analysis. The data is shown in Figure 5, and Figure 6 shows the fitted spectrum together with a simulated spectrum based on the expectation for the same structure if the isotopes had been terrestrial. The quality of the 2069 isotope fit in Table 4 is excellent, with the positive satellites matching data to better than 2%. The negative ones, being closer to noise, achieve a fit to better than 10% in amplitude. Other lithium-bearing structures are listed in S2 and their relationship to the primary structure at 2320 is mapped in S4.

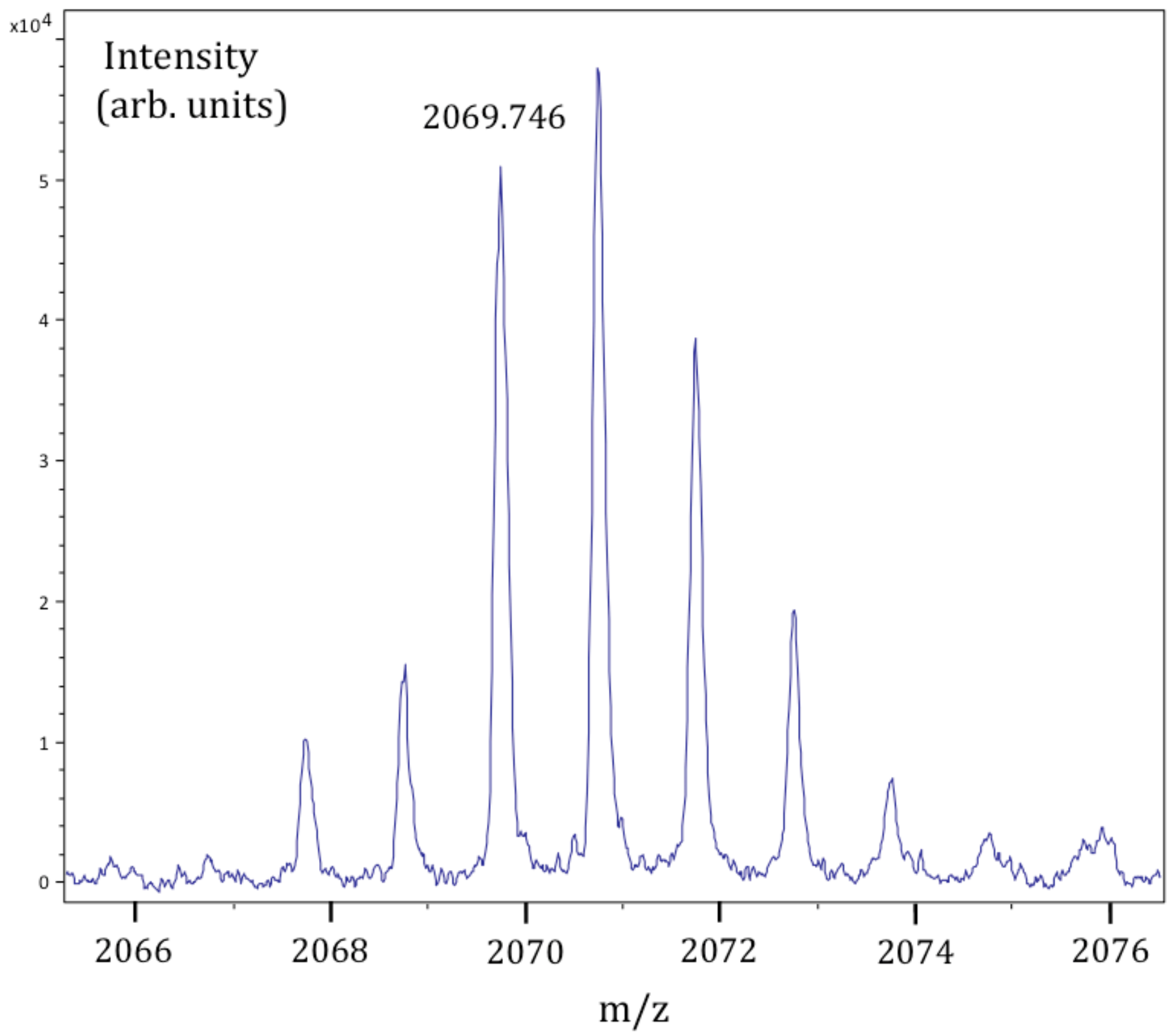

**Figure 5. Spectrum in the region of m/z 2069. Phase P2, matrix SA.**

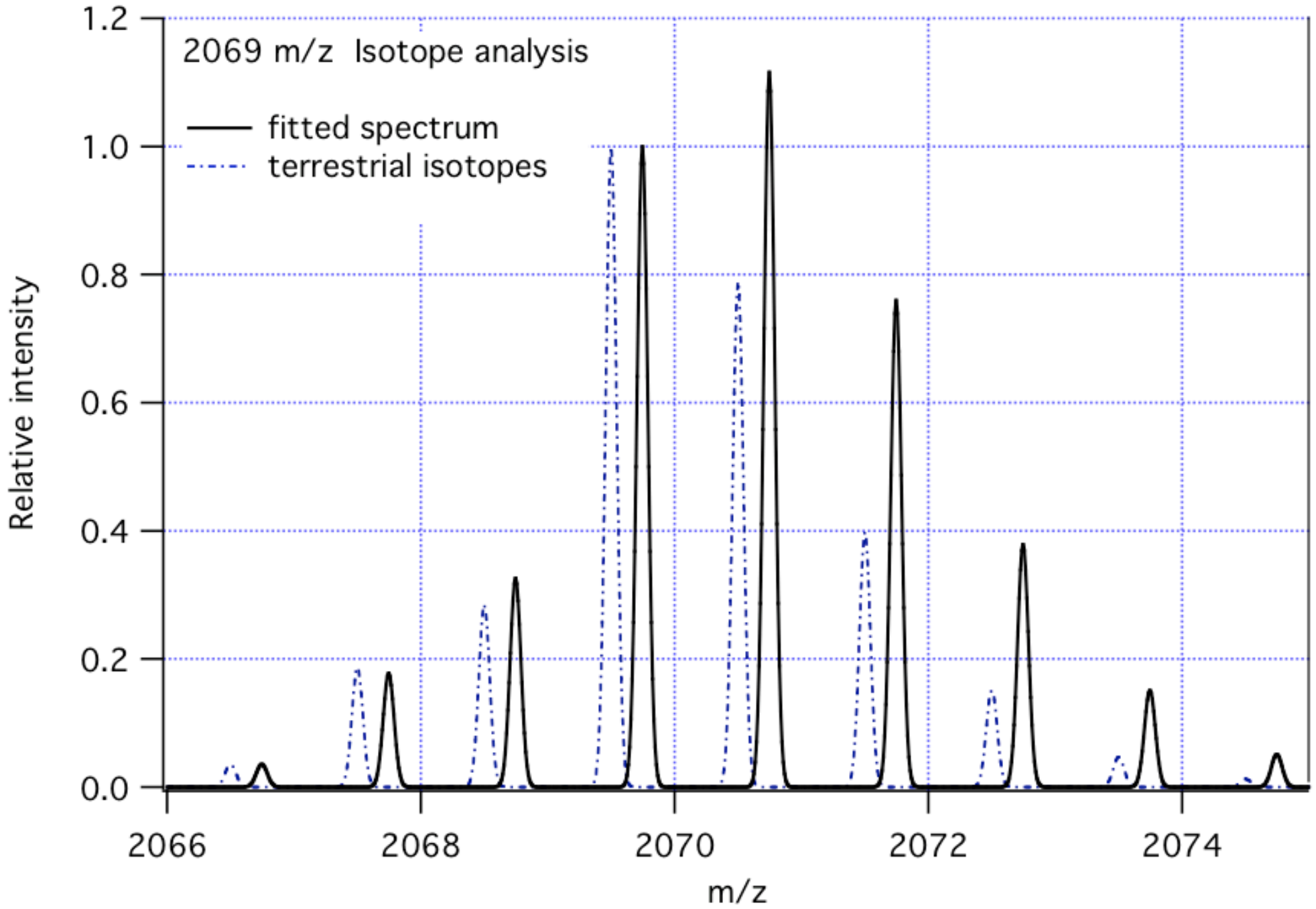

**Figure 6. Fitted isotope spectrum (solid line) to 2069 peak and simulation of the same structure with terrestrial isotope values (dotted line, offset for clarity). Details of the fit are given in Table 4.**

**Table 4. Isotope satellite analysis for peak at m/z 2069 of Figure 5. S/N = 63. Modeled structure = (26Gly, 4Gly$_{OH}$, 3Fe) + 7'O' + 2Li + H.**

| **Satellite** | **(-3)** | **(-2)** | **(-1)** | **(0)** | **(1)** | **(2)** | **(3)** |
|---|---|---|---|---|---|---|---|
| Data | 1,992 | 10,151 | 15,465 | 50,924 | 57,957 | 38,699 | 19,170 |
| Fitted $^2$H 25,000 ‰<br>Preset $^{15}$N 1,015 ‰ | .009 | .044 | .081 | .248 | .277 | .189 | .094 |
| Normalized data | .010 | .049 | .075 | .248 | .282 | .188 | .093 |
| Calc. terrestrial | .009 | .046 | .070 | .248 | .196 | .099 | .038 |

**Achieving a unified structure via "MS/MS" and other analyses.**

It is of the greatest interest to establish whether there is a single molecular specie giving rise to all of the observed peaks in these mass spectra. There could be one, two, ten or fifty structures in nature composed from glycine, iron, oxygen, lithium and hydrogen. However, time and again, a relatively simple and consistent mass spectrum is observed. The spectra are in part matrix-dependent because each matrix has different ionization properties and donates protons with different speeds that vary by orders of magnitude [10]. In the case of the present data we see partial fragmentation during the MALDI process of a very few pre-existing molecular entities. The MS/MS measurements were a powerful tool in structural analysis that helped establishing similarities and differences in fragmentation patterns of different parent ions, as shown in S4.

In the present study we had sufficient signal level to perform an MS/MS analysis on signals at m/z 1567, 1639 and 2124. These were all accomplished with CHCA as the matrix owing to its more specific definition of an m/z peak than is achievable with SA. With the molecule under study the SA matrix had a more gentle action and did not strip a molecular fragment free of loose protons, etc. so thoroughly as did CHCA. In addition, the SA matrix readily complexed with the molecule under study, presenting difficulties of interpretation. In general, any candidate peak in the SA spectrum had to be tested for a "root" peak at m/z 223 less, owing to the prevalent complexing of ($M_{SA}$-H) = 223 with the molecule under study. Peaks that showed this difference were rejected from consideration, being matrix artifacts.

The most productive 1567 MS/MS spectrum, obtained with CHCA, is displayed in S3, together with consistent structures for all of its fragments. Each fragment has to appear within the applicable MS/MS "root" molecular structure and indeed the process begins with analysis of the fragments and works back to a root structure. Further than this, the root structures for signals at m/z 1567 and 1639 are found to be consistent with each other (1567 is contained within the 1639 MS/MS spectrum) and also consistent with the higher 2124 root structure that can now be constructed with a fair degree of confidence. These structures are related to each other in the "tree" of S4. In that drawing the MS/MS series for 1567 and 1639 are displayed in the left two columns, a general spectrum from 2124 downwards is displayed in the third column, and the last column collects the most intense lithium-bearing fragments. Where signal-to noise ratios allowed, the displayed species in Figure S4.1 were the subject of the isotope analysis described above, with results listed in S2.

Also taken into consideration in the determination of the molecular tree was the prior observation of 4641Da molecules from the same preparation [2], which now can be identified as adducts or dimers of the 2320 molecule, an assignment presented in more detail in S5. It is concluded that two or three "master" structures around m/z 2320 are the most probable source for all of the fragments in a) first pass MALDI with three different matrices (CHCA, SA and DHB); b) MS/MS of the three peaks with sufficient amplitude for second pass analysis and c) the components of the 4641Da cluster of peaks. The candidate "master" structures are, in the notation (i,j,k) = (Gly, $Gly_{OH}$, Fe):
2320 = (26,6,4) + 10'O' + 2Li + 2H
2364 = (24,8,4) + 11'O' + Li + 5H
2124 = (26,4,4) + 8'O' -2H
Structures are presented in S5 for each of these. 2320 and 2364 have 32 amino acid residues while 2124 has 30. An additional 34 amino acid structure at 2402 is also proposed in S5, to fit two peaks within the 4641Da cluster. The chain lengths in each anti-parallel configuration are therefore 16 residues for 2320 and 2364, 15 residues for 2124 and 17 residues for 2402. As discussed herein the glycine residues are subject to varying degrees of hydroxylation, but the more constant parameter is the number of glycines per side. We propose calling the 16-residue entity Hemolithin16, and the 15 or 17-residue entities Hemolithins 15 or 17. Molecular modeling of the 2320 entity using Spartan software [11] with the MMFF force field [12] is shown in Figure 7.

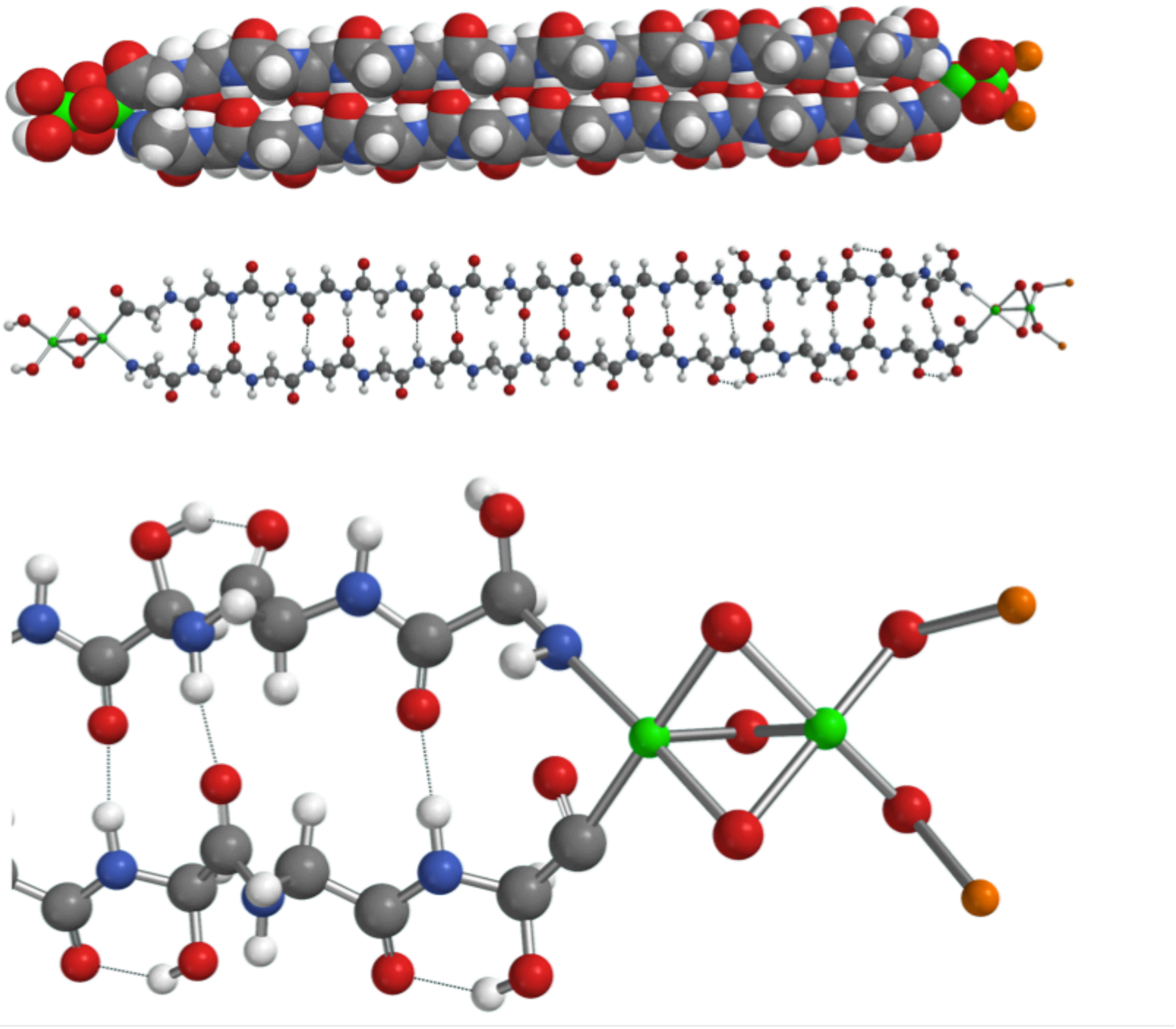

**Figure 7 Model of the 2320 hemolithin molecule after MMFF energy minimization. Top: in space-filling mode; Center: ball and stick; Bottom: enlarged view of iron, oxygen and lithium termination. White = H; orange = Li; grey = C; blue = N; red = O and green = Fe. Hydrogen bonds are shown by dotted lines.**

## DISCUSSION

Progress toward this structure has only been possible due to higher signal-to-noise data that allowed us to quantify the number of iron and lithium atoms associated with a mass spectrum peak, via their satellite amplitudes in the (-1) or (-2) positions relative to the intensity of the (0) and (+1) peaks. A program was written (S1) to include all the stable isotopes of H, Li, C, N, O and Fe and calculate the full isotope spectrum for any trial compound containing them. Complex isotope spectra could be fitted to within a few percent in many cases, data summarized in S2. Starting with the clear presence of polymers of glycine and hydroxy-glycine both in the present lower m/z spectra and in the prior results [1,2], and adding the newly-determined numbers of iron and/or lithium, it was found that only additional oxygen atoms, sometimes in large numbers, were needed to match an unknown m/z peak. The details of the smaller fragments in the 1567 MS/MS spectra (S3) enabled us to piece together the $FeO_3Fe$ motif so prevalent throughout the data. Trial structures were simulated using the "Spartan" software from Wavefunction [11,12] in MMFF and at higher levels. The glycine strands gave a lower energy when connected by hydrogen bonds in anti-parallel, as opposed to parallel, helping to decide a structural issue that was not apparent from the mass spectra. In a conceptual breakthrough iron, with its five valence bonds radiating outward, was considered for the role of terminating the strands, with a peptide 'C' terminus and an 'N' terminus joining onto one iron atom. The remaining three valencies could carry oxygen atoms that themselves bonded to a second iron atom. Such configurations when modeled in MMFF were extremely quick to converge, remaining exactly straight when iron closed out each end of the polymer. The straight 2320 molecule is shown in space-filling format in Figure 7 (top) and schematically in Figure 7 (center and bottom). Its length is 7nm.

In a review of the known binding geometries for iron within proteins such as hemoglobin or hemerythrin, this new motif was not found. Its overwhelming simplicity and the new association with beta sheet proteins make its future study very important.

In agreement with two prior MALDI studies of polymer amide in meteorites [1,2] the spectral peaks had higher (1,2,3 etc.) satellite intensities than would be calculated for the same structures using only terrestrial isotope values. In fact, from fitting the spectra, the average enhancement of $^{2}H$ above terrestrial was found to be 25,700‰, with a standard deviation of 3,500‰, n=15 (S2). This may also be expressed as a ratio $D/H = (4.1 \pm 0.5) \times 10^{-3}$. It was not found necessary to vary Li, C, O or Fe isotope ratios away from terrestrial in order to fit the data. Only $^{15}N$ is known to be raised in this same Acfer 086 meteorite sample [9], by a completely different technique, and its enhancement value of 1,015 ± 220 ‰ was incorporated as a baseline. The $^{2}H$ enhancement quoted here is calculated with this $^{15}N$ enhancement as a preset condition of the fit to data. If desired, the $^{15}N$ value can be changed in the future with predictable effect on the $^{2}H$ determination. If we consider only polymer glycine with its H:N ratio of 3:1, a change to N(‰) of -1,000 would be equivalent to a change to H(‰) of +7,700, approximately, in the fit to data.

Such high $^{2}H$ (D) enhancements are well documented in molecular clouds [13] where low-temperature isotopic selection has operated for timespans of more than 10 million years. D/H fractions in simple molecules within the clouds are generally in the region of 0.01 to 0.1, to be compared with terrestrial $D/H = 1.55 \times 10^{-4}$. Our measured average enhancement of 25,700‰ is equivalent to a D/H ratio of 0.0041, which would be consistent with an interstellar origin for the hemolithin molecule. We have estimated [14] that simple amino acids should be able to slowly polymerize in the conditions of "warm dense molecular clouds", without necessarily requiring surfaces for the reaction. Furthermore, the elements H, Li, C, N, O and Fe comprising this molecule were initially the most abundant when the first massive stars released them about 13B years ago.

Deuterium is also enriched in molecules of the proto-planetary disc. Aikawa and Herbst [15] have applied complete molecular cloud chemistry models to the disc and one of their predictions relates to the amount of DCN/HCN enrichment expected at 30AU radius, which is the likely region for the origin of comets. There they found that DCN/HCN converged in calculations to a value of 0.004, only slightly larger than the ratio measured in comet Hale-Bopp [16,17] of $(2.3\pm0.4)x10^{-3}$. The D/H ratio of 0.0041 that we find in polymer amide is therefore consistent with both calculated and measured DCN/HCN ratios for comets and outer solar system material. Further to this, the $^{15}N$ enhancement that we measured by a different technique is also consistent with cometary values [9]. Despite this consistency, the presumed source of CV3 class meteorites containing hemolithin is a parent body in the asteroid belt that is only between 2.2 and 3.3AU from the sun.

High deuterium enhancement, comparable to the present readings, has been found in ultra-carbonaceous micro-meteorites from Antarctic snow [18]. The method of detection was secondary ion mass spectrometry (SIMS). There was no sign of these particles having experienced high temperature and because found in snow, rather than ice, there was less chance of aqueous alteration. The highest D/H reading was $(4.6\pm0.5)x10^{-3}$, comparable to our present average of $(4.1 \pm 0.5)x10^{-3}$ for the meteoritic protein. The polyglycine chains of the present work have C/H = 0.66, which is rather lower than reported in [18] where C/H ratios were generally greater than 1 and indeed the organic carrier of the deuterium was not identified. In [18] the presence of crystalline silicates as opposed to amorphous silicates in small inclusions was cited as evidence in favor of a protoplanetary rather than an interstellar origin, although different ultimate origins could have applied to the organic and mineral regions. As in the present work there was minimal processing before measurement, which could be important in the determination of deuterium.

It appears that three common variants of Hemolithin with chain lengths 15, 16 and 17 may co-exist. If this molecular type grows spontaneously when its glycine, Fe, O, Li and H components are present together in a particular environment then it is difficult to imagine such a tight length distribution being the result. A function yet to be identified could determine the length and maintain such a tight distribution. If successful in that function, energy could be harvested, which is a thermodynamic requirement [19], to aid in the creation of molecular copies of identical length.

Although not directly related to the present finding, polymers of up to 16 glycine residues have been catalytically assembled on a $TiO_2$ surface [21], out of the gas phase. The distribution of polymers was "diffusion-like" with a maximum intensity in mass spectrometry at around 5 glycine residues in length. In [21] there also was evidence of a degree of beta sheet organization after hydration. In our own work the possibility of "in vitro" glycine polymerization can be discounted because:
a) The distribution of polymer lengths in the "master" structures is very tight around 16 residues,
b) Extended polymer amide -C-C-N- backbone chains existed within the intact dry meteorite prior to solvent extraction as evidenced by the exact 2:1 ratio of CN to $C_2$ negative ions in focused ion beam (FIB) SIMS milling [9], associated with high $^{15}N$ levels in the CN ions.
c) The solvent extraction time was only 24 hours at room temperature (approximately 293K), whereas 403K had been used in the surface-catalyzed case [21].
d) Terrestrial contamination glycine was not involved because of the very high extra-terrestrial D/H ratios.

The $FeO_3Fe$ motif at the ends of hemolithin has been studied in relation to the photo-induced splitting of water [20] on hematite surfaces, pointing to a direction for further investigation. As a footnote, the outdated assignments in Table 2 are explained by the mass equivalence of (2 x alanine + t=18) to $FeO_3Fe$, each having mass 160, and similar items. In the prior work [1,2] the signal-to-noise ratios were too low to identify Fe and Li via their light isotopes, although Li was deduced to be present simply via its dominant mass 7 isotope [2].

## CONCLUSIONS

One-step room temperature solvent extraction from micron scale particles of the meteorite Acfer 086 (class CV3), has yielded via MALDI mass spectrometry a relatively simple spectrum in the m/z range 1-2,400 that is dominated by a single protein type. This is composed of anti-parallel beta strands of glycine, each of 15 to 17 -residue length, with about 20% oxidation to hydroxy-glycine, and termination at each end with an iron atom directly bonded to C and N terminals of the peptide strands. There are additional tri-oxygen/iron groups at each end, and lithium adducts. The principal indicator of extra-terrestrial origin is an extreme raised D/H ratio that is revealed by close quantitative fitting of isotopic satellite peaks. The average molecular deuterium excess above terrestrial is (25,700 ± 3,500)‰, or a D/H ratio of $(4.1 \pm 0.5) \times 10^{-3}$, comparable to cometary levels, interstellar levels and also equal to the highest prior report in micro-meteorites. The iron oxide grouping at the tips of the molecule is of a type studied in other contexts for the photo-splitting of water.

## ACKNOWLEDGEMENTS

Authors MWM and JEMM wish to thank Professor Guido Guidotti of Harvard for encouragement and advice and Brian Stall at Bruker for allowing the use of their instrument for this analysis.

# Supplementary Information
**S 1. Theory of isotopic satellites**

The less commonly observed (-1) and (-2) satellites to a principal m/z peak are used to find the number of iron and lithium atoms in any given m/z fragment of the protein under study. Conventional amino acid polymers that only contain H, C, N, O and S atoms, all display "heavy isotope" peaks to the higher mass side of the principal peak giving predictable (+1), (+2) etc. amplitudes once the composition of a molecule is known. The "light" iron and lithium satellites to the low mass side of a peak fortunately differ from each other, with $^{54}Fe$ being at (-2) and $^{6}Li$ at (-1). The (-1) and (-2) amplitudes are therefore clear indicators of iron and lithium content, although there are higher order interactions that have to be included in the fitting. The object here is to test the fidelity of a satellite spectrum against the terrestrial expectation and if necessary adjust iron, lithium and one or more of the heavy isotope ratios (particularly $^{2}H$) to trial extra-terrestrial levels until a spectrum is accurately reproduced. In this section the mechanism for this calculation is described.

We start with terrestrial heavy isotope fractions ([22] Vienna) as reference values:

| | |
|---|---|
| VSMOW water | $R_H = {}^{2}H/{}^{1}H = 155.76 \pm 0.05 \times 10^{-6}$ |
| VSMOW water | $R_O = {}^{18}O/{}^{16}O = 2{,}005.20 \pm 0.45 \times 10^{-6}$ |
| V-PDB | $R_C = {}^{13}C/{}^{12}C = 11{,}237.2 \times 10^{-6}$ |
| Atmospheric Nitrogen | $R_N = {}^{15}N/{}^{14}N = 3{,}612 \pm 7 \times 10^{-6}$ , |

to which we add the iron and lithium terrestrial values [8]:

| | |
|---|---|
| Iron | $R_{F54} = {}^{54}Fe/{}^{56}Fe = 0.06371$ |
| | $R_{F57} = {}^{57}Fe/{}^{56}Fe = 0.02309$ |
| Lithium | $R_{L6} = {}^{6}Li/{}^{7}Li = 0.08217$ |

$^{58}Fe$ is at the 0.2% level and is neglected here.

We first illustrate the calculation of isotope probabilities with reference to $^{1}H$ and $^{2}H$ and then extend to the full array. In a molecule with $n$ hydrogen ($^{1}H + {}^{2}H$) atoms in its formula we calculate the probability of obtaining $k$ $^{2}H$ atoms from the binomial distribution as

$$P_H(k) = \frac{n!}{k!(n-k)!} p_H^k (1-p_H)^{n-k}$$

where $p_H = \frac{[^{2}H]}{[^{1}H]+[^{2}H]}$ is the probability of a heavy isotope substitution. For example, if the ratio of heavy to light isotopes is $R_H = {}^{2}H/{}^{1}H$ then $p_H = R_H/(1+R_H)$ = 1.5574 x $10^{-4}$ for the terrestrial hydrogen standard. Similar considerations lead to values for $p_C$, $p_N$, $p_O$, $p_{F54}$, $p_{F57}$ and $p_{L6}$ denoted in general by $p_X$ in Table S1.1.

We illustrate the calculation with a terrestrial matrix cluster [7] at m/z=861 that is frequently observed in MALDI spectra with matrix CHCA. The 861 molecule is due to ($M_4KNa_3 - 3H$) in which $M$ is the CHCA molecule of mass 189.043 with formula $M = C_{10}H_7NO_3$. We will calculate its isotope satellites and compare them with the observed matrix cluster (Figure S1.1, detail from [1], Fig. 2) which has isotopic (Δm = +1 and Δm = +2) satellites of amplitude 47% and 17% relative to the main peak.

The composition of the '861' cluster is $H_{25}C_{40}N_4O_{12}K_1Na_3$. Of these constituents only $^{23}Na$ is naturally mono-isotopic. In the following table its heavy isotope probability is zero. Entries in the table are obtained from the binomial distribution above for $k$ = 0,1,2,3 heavy substitutions in the $n$ atoms of the species concerned, with $n$=25 for hydrogen, etc.

**Table S1.1 Isotope satellite calculation for the '861' matrix cluster peak**

| | $H_{25}$ | $C_{40}$ | $N_4$ | $O_{12}$ | $K_1$ | $Na_3$ |
|---|---|---|---|---|---|---|
| $p_X$ | 1.5574 x $10^{-4}$ | 0.011112 | 3.599 x $10^{-3}$ | 2.0012 x $10^{-3}$ | 0.06731 | 0 |
| $P_X(0)$ | 0.996 | 0.640 | 0.986 | 0.976 | 0.933 | 1 |
| $P_X(1)$ | 0.0039 | 0.288 | 0.0036 | 0.023 (Δm = +2) | 0.067 (Δm = +2 | 0 |
| $P_X(2)$ | 7 x $10^{-6}$ | 0.063 | - | - | - | - |
| $P_X(3)$ | - | 0.009 | - | - | - | - |

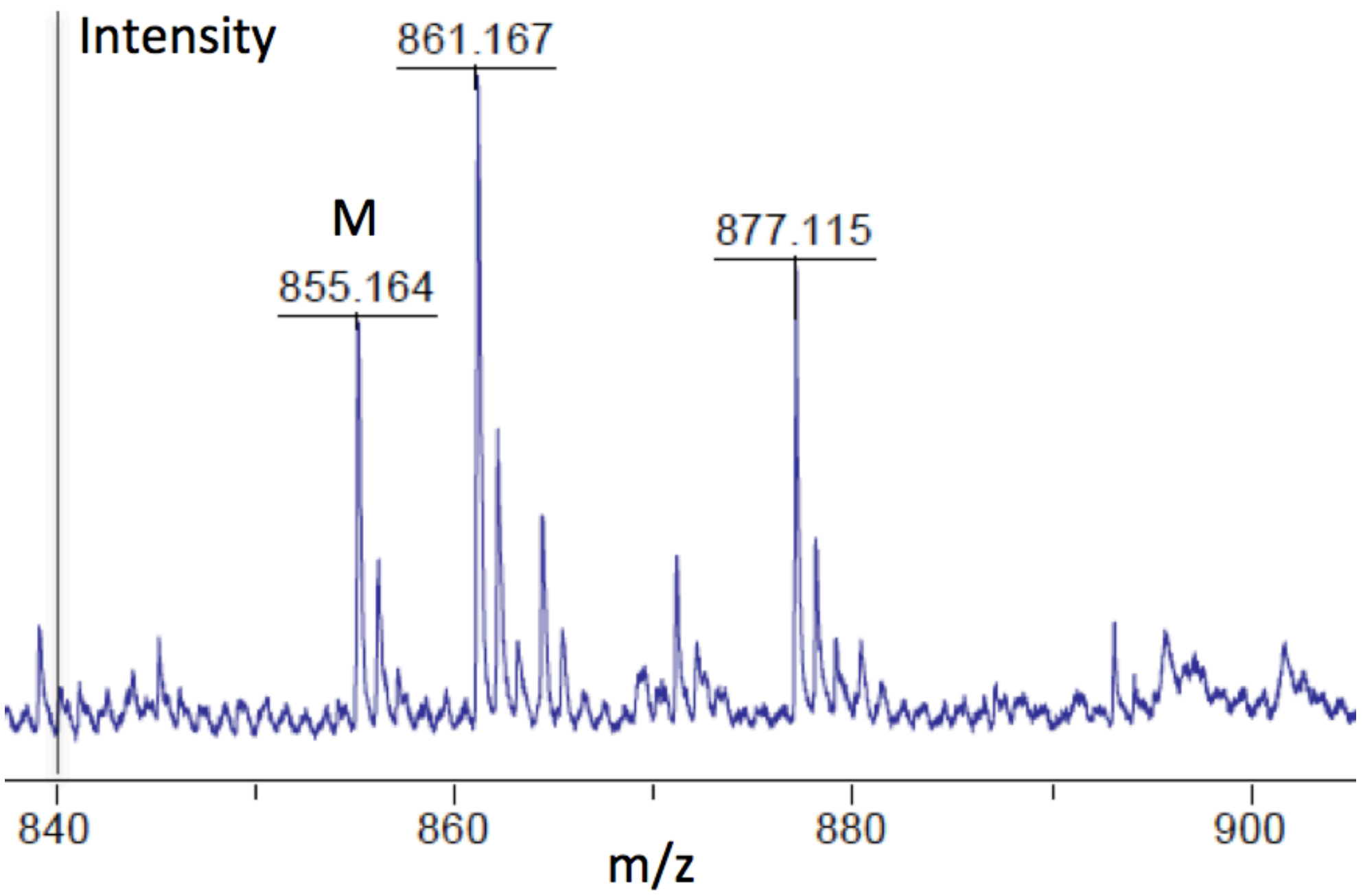

**Fig S1.1. Matrix cluster peaks at m/z 855, 861 and 877 reproduced from [1], Fig. 2.**

The isotope ratios are presumed uncorrelated, so the probability of having zero mass addition is the product of the individual $P_X(0)$ terms:
$\Pi(0) = P_H(0)P_C(0)P_N(0)P_O(0)P_K(0)P_{Na}(0) = 0.572$

There are three different ways to have Δm = +1, involving $H$, $C$ and $N$ respectively. Oxygen and potassium jump to Δm = +2 immediately. The probability of having Δm = +1 is therefore:

$$\Sigma\, \Pi(1) = P_H(1)P_C(0)P_N(0)P_O(0)P_K(0)P_{Na}(0) + P_H(0)P_C(1)P_N(0)P_O(0)P_K(0)P_{Na}(0) + P_H(0)P_C(0)P_N(1)P_O(0)P_K(0)P_{Na}(0) = 0.262$$

Similarly, the possible combinations that give $\Delta m = +2$ sum to give $\Sigma\, \Pi(2) = 0.113$ and the $\Delta m = +3$ sum is $\Sigma\, \Pi(3) = 0.033$. By now, we have accounted for 0.98 of the collective isotope amplitude, with the remaining 0.02 assigned to $\Delta m = +4$ and higher species. The total of probabilities should sum to unity, when correctly carried to completion.

Continuing with our terrestrial example at m/z 861 we normalize the $\Delta m = 0$ peak to 100%, as listed in Table S1.2. In the calculated satellites of the 861 matrix peak there is good agreement with the $\Delta m = +1$ and $\Delta m = +2$ measured intensities but the $\Delta m = +3$ location appears to also carry another peak that obscures the true reading.

**Table S1.2 Calculated and experimental isotope satellite ratios for the 861 matrix cluster peak.**

| | Δm = 0 | Δm = +1 | Δm = +2 | Δm = +3 |
|---|---|---|---|---|
| Calc. | 100% | 45.8% | 19.8% | 5.8% |
| Expt. | 100% | 47% | 17% | - |

We handle the complexity of the present calculations via a new computer program that inputs a trial molecular composition and trial values for the isotope ratios, whether heavy or light. The output comprises peak amplitudes from (-6) to (+6) that can then be compared with data. We found that it was efficient to start the comparison with a trial $^{2}H$ enhancement of 25,000‰, on top of the preset $^{15}N$ enhancement of 1,015‰ discussed in the main text. Depending upon the closeness of the fit the $^{2}$H enhancement was adjusted in increments of ± 2,500‰ until the various satellite peaks were collectively fitted. If the (-2) and (-1) peaks did not clearly indicate a number of iron or lithium atoms, either the formula under test was adjusted, or the fitting attempt was abandoned for that peak because of inadequate signal-to-noise ratio. After a formula adjustment the whole set of satellites has to be fitted again because of the numerous “knock-on” effects from one satellite to another.

## S2. Data Compilation.

**Table S2.1 Isotope fitting to peaks for given structures in notation (i,j,k) = (Gly, $Gly_{OH}$, Fe); *m*'O' represents oxygen atoms associated with iron. $^2H/^{15}N$ per mil enhancements from fit to data with preset $\delta^{15}N$ = 1,015‰ [9]. Average $^2H$ enhancement = 25,700‰ (σ = 3,500‰, n = 15) excluding 1417.**

| Obs. m/z<br>Integer m/z | Satellite | | | | | | | S/N<br>0 peak |
|---|---|---|---|---|---|---|---|---|
| 2364.385 | (-3) | (-2) | (-1) | 0 | (1) | (2) | (3) | |
| Data | .012 | .053 | .074 | .220 | .228 | .195 | .100 | 16 |
| Fit | .009 | .048 | .076 | .220 | .264 | .197 | .108 | |
| **2364** | (24,8,4) + 11'O' + Li + 5H 22,500/1,015 | | | | | | | |
| | | | | | | | | |
| 2124.885 | (-3) | (-2) | (-1) | 0 | (1) | (2) | (3) | |
| Data | .008 | .045 | .065 | .237 | .279 | .202 | .105 | 135 |
| Fit | .006 | .053 | .067 | .237 | .278 | .195 | .099 | |
| **2124** | (26,4,4) + 8'O' -2H 25,000/1,015 | | | | | | | |
| | | | | | | | | |
| 2069.746 | (-3) | (-2) | (-1) | 0 | (1) | (2) | (3) | |
| Data | .010 | .049 | .075 | .248 | .282 | .188 | .093 | 63 |
| Fit | .009 | .044 | .081 | .248 | .277 | .189 | .094 | |
| **2069** | (26,4,3) + 7'O' +2Li + H 25,000/1,015 | | | | | | | |
| | | | | | | | | |
| 1866.804 | (-3) | (-2) | (-1) | 0 | (1) | (2) | (3) | |
| Data | .030 | .053 | .093 | .274 | .332 | .185 | .101 | 27 |
| Fit | .006 | .049 | .072 | .274 | .293 | .187 | .087 | |
| **1866** | (24,4,2) + 5'O' + 2Li 25,000/1,015 | | | | | | | |
| | | | | | | | | |
| 1639.610 | (-3) | (-2) | (-1) | 0 | (1) | (2) | (3) | |
| Data | .013 | .075 | .071 | .300 | .291 | .165 | .069 | 112 |
| Fit | .007 | .070 | .071 | .300 | .287 | .162 | .067 | |
| **1639** | (20,2,4) + 8'O' + H 27,500/1,015 | | | | | | | |
| | | | | | | | | |
| 1567.678 | (-3) | (-2) | (-1) | 0 | (1) | (2) | (3) | |
| Data | .007 | .055 | .051 | .314 | .318 | .164 | .067 | 135 |
| Fit | .004 | .060 | .060 | .314 | .301 | .168 | .068 | |
| **1567** | (20,2,3) + 7'O' + H 27,500/1,015 | | | | | | | |
| | | | | | | | | |
| 1417.606 | (-3) | (-2) | (-1) | 0 | (1) | (2) | (3) | |
| Data | - | 0 | 0 | .261 | .311 | .233 | .129 | 26 |
| Fit | 0 | 0 | 0 | .261 | .338 | .231 | .111 | |
| **1417** | (21,3,0) + H 50,000/1,015 | | | | | | | |
| | | | | | | | | |
| 1381.534 | (-3) | (-2) | (-1) | 0 | (1) | (2) | (3) | |
| Data | .017 | .076 | .107 | .335 | .249 | .147 | .073 | 13 |
| Fit | .012 | .062 | .098 | .335 | .279 | .139 | .051 | |
| **1381** | (17,2,3) + 5'O' + 2Li + 4H 30,000/1,015 | | | | | | | |
| | | | | | | | | |
| 1226.527 | (-3) | (-2) | (-1) | 0 | (1) | (2) | (3) | |
| Data | .001 | .055 | .034 | .402 | .309 | .147 | .039 | 36 |
| Fit | .001 | .049 | .038 | .402 | .307 | .138 | .046 | |
| **1226** | (11,6,2) + 3'O' + H 27,500/1,015 | | | | | | | |
| | | | | | | | | |
| 1049.474 | (-3) | (-2) | (-1) | 0 | (1) | (2) | (3) | |
| Data | .011 | .054 | .056 | .432 | .310 | .137 | .051 | 31 |
| Fit | .005 | .054 | .068 | .432 | .284 | .113 | .033 | |
| **1049** | (11,3,2) + 5'O' + Li + 4H 30,000/1,015 | | | | | | | |
| | | | | | | | | |
| 1043.344 | (-3) | (-2) | (-1) | 0 | (1) | (2) | (3) | |
| Data | .016 | .070 | .097 | .423 | .275 | .114 | .037 | 114 |
| Fit | .009 | .055 | .096 | .423 | .272 | .104 | .030 | |
| **1043** | (15,0,2) + 4'O' + 2Li – 2H 27,500/1,015 | | | | | | | |
| | | | | | | | | |
| 1027.253 | (-3) | (-2) | (-1) | 0 | (1) | (2) | (3) | |
| Data | - | .032 | .023 | .470 | .320 | .125 | .035 | 111 |
| Fit | 0 | .029 | .020 | .470 | .311 | .123 | .036 | |
| **1027** | (11,4,1) + 3'O' + 4H 25,000/1,015 | | | | | | | |
| | | | | | | | | |
| 765.350 | (-3) | (-2) | (-1) | 0 | (1) | (2) | (3) | |
| Data | - | .048 | .152 | .517 | .225 | .069 | .020 | 24 |
| Fit | - | .042 | .132 | .517 | .222 | .063 | .013 | |
| **765** | (10,1,1) + 3'O' + 3Li -3H 20,000/1,015 | | | | | | | |
| | | | | | | | | |
| 730.306 | (-3) | (-2) | (-1) | 0 | (1) | (2) | (3) | |
| Data | .008 | .038 | .025 | .572 | .262 | .083 | .014 | 84 |
| Fit | 0 | .036 | .017 | .572 | .273 | .080 | .018 | |
| **730** | (8,3,1) –H 25,000/1,015 | | | | | | | |
| | | | | | | | | |
| 531.335 | (-3) | (-2) | (-1) | 0 | (1) | (2) | (3) | 33 |
| Data | - | - | .003 | .691 | .243 | .054 | .098 | |
| Fit | 0 | 0 | 0 | .691 | .243 | .055 | .095 | |
| **531** | (9,0,0) + 'O' + 2H 17,500/1,015 | | | | | | | |
| | | | | | | | | |
| 401.086 | (-3) | (-2) | (-1) | 0 | (1) | (2) | (3) | |
| Data | - | .010 | .005 | .717 | .217 | .061 | .015 | 182 |
| Fit | 0 | 0 | 0 | .717 | .229 | .046 | .007 | |
| **401** | (7,0,0) + 2H 30,000/1,015 | | | | | | | |

## S3. Structural synthesis

## S3.1 Structural synthesis via the MS/MS spectrum at m/z = 1567

The 1567 peak that appeared prominently in the full spectrum (Figure 3, main paper) contained upon MS/MS analysis more than 29 peaks at signal-to-noise ratio between 3 and 35, as shown in Figures S3.1 and S3.2. The highest intensity peaks were analyzed to yield the 27 structures of fragments consistent with an origin in the 1567 molecule that are listed in Table S3.1. Peaks with signal-to-noise greater than 20 were subject to isotope analysis. The remaining peaks were constructed to fit the observed mass while remaining consistent with the master 1567 structure given at the top of Table S3.1 .

There is support in this MS/MS analysis for many of the fundamental features of hemolithin:
a) polymers of glycine and hydroxy-glycine appear throughout, often attached to iron atoms, but also on their own in the mass range from 531 to 677.
b) many oxygen atoms are contained in hemolithin, in three “reservoirs”:
R1 as part of each peptide residue at the carbonyl group – relatively stable;
R2 as part of hydroxy-glycine, which is marginally stable;
R3 as part of the two-atom iron complex which, again, is relatively stable.
c) A residue count of 22 exists in 1567, but the 1080 and 1009 fragments suggest a peptide chain length of 15 residues between iron atoms. The 1080 fragment definitely has two Fe atoms according to its isotope analysis.

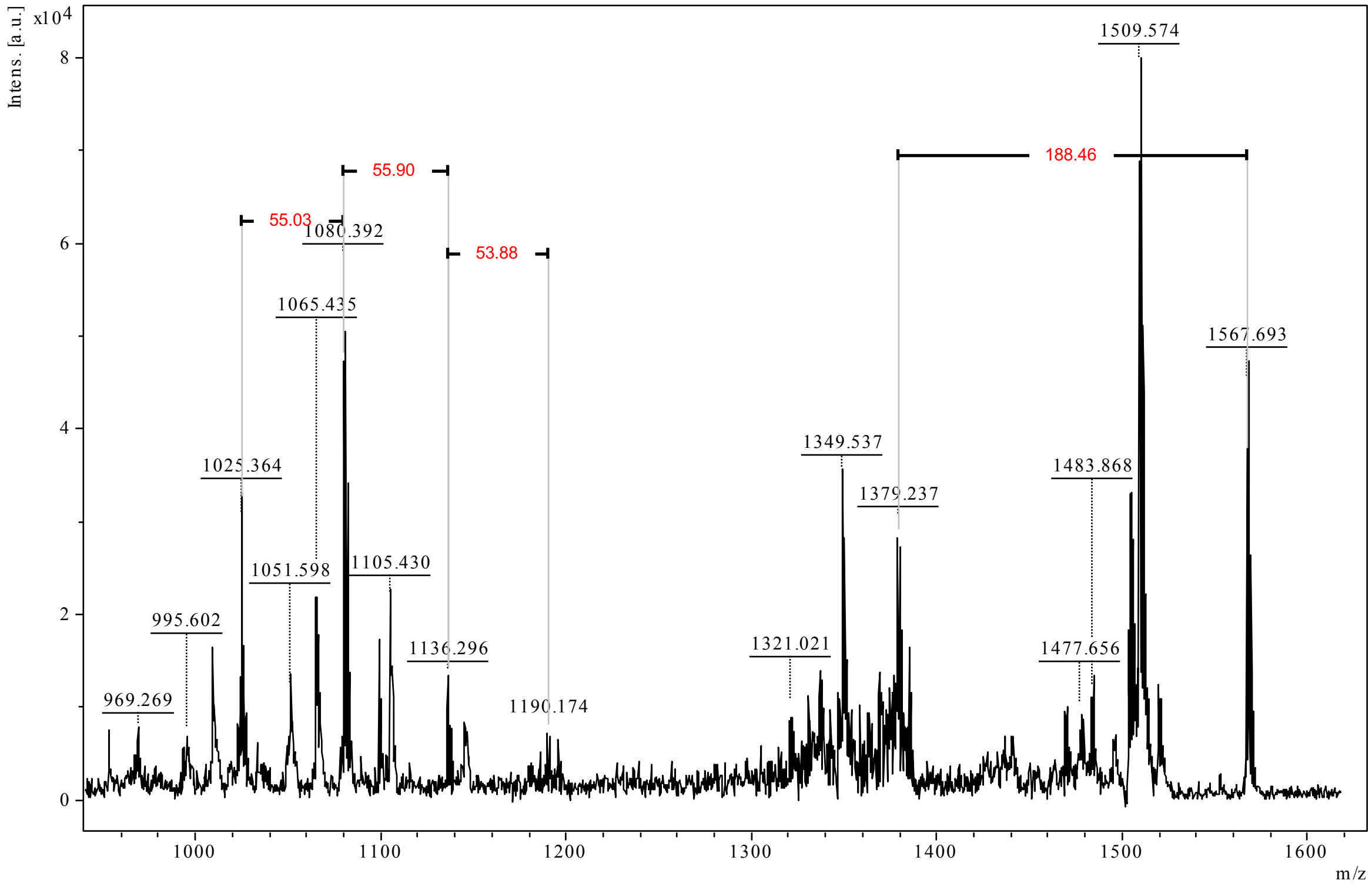

**Figure S3.1 The 1567 MS/MS spectrum in the m/z range 1000 to 1600.**

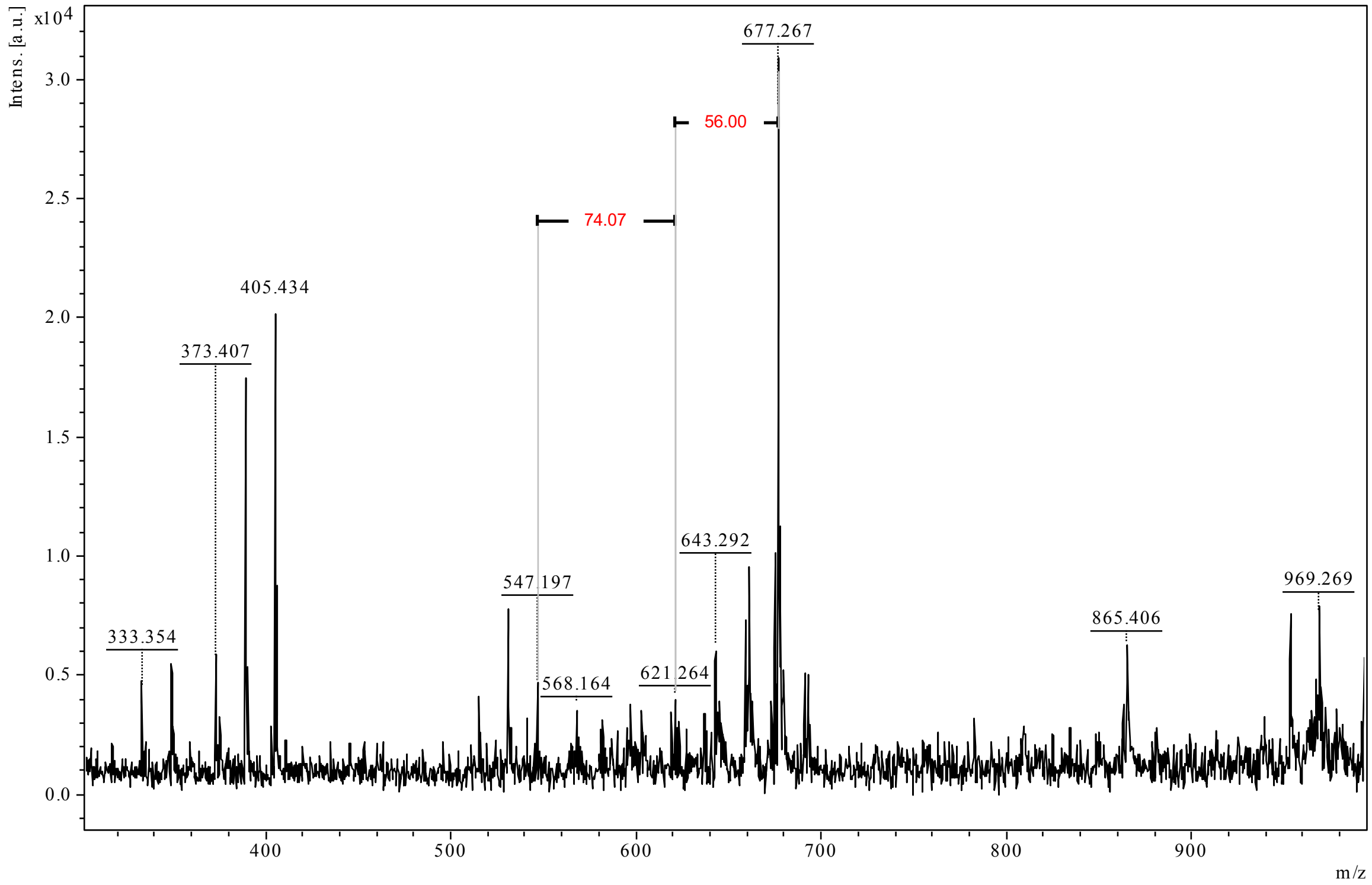

**Figure S3.2 The 1567 MS/MS spectrum in the m/z range 300 to 1000**

**Table S3.1 Listing of peaks in the 1567 MS/MS spectrum.**

| Observed m/z | Intensity | Structure | S/N |
|---|---|---|---|
| 1567.691 | 241,061 | 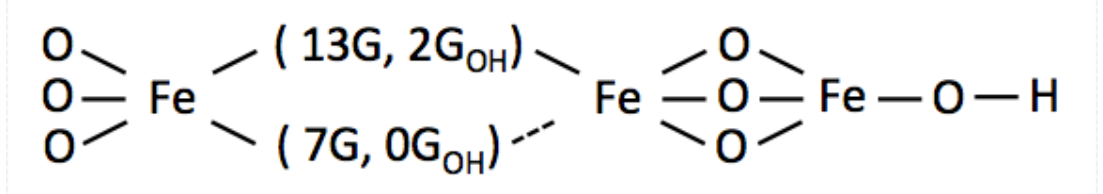 | 20 |
| 1519.921 | 12,433 | 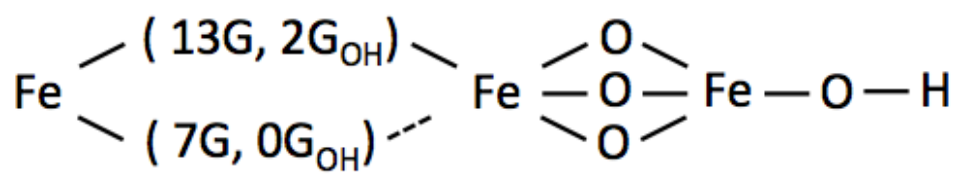 | 6 |
| 1509.574 | 68,734 | 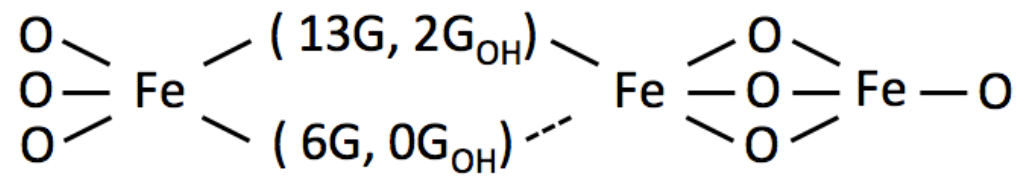 | 35 |
| 1503.819 | 18,359 | 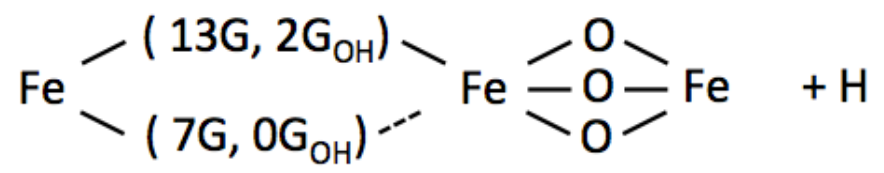 | 15 |
| 1469.426 | 9,572 | 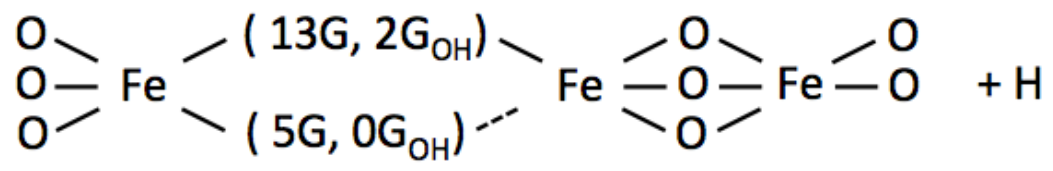 | 3 |
| 1379.237 | 28,185 | 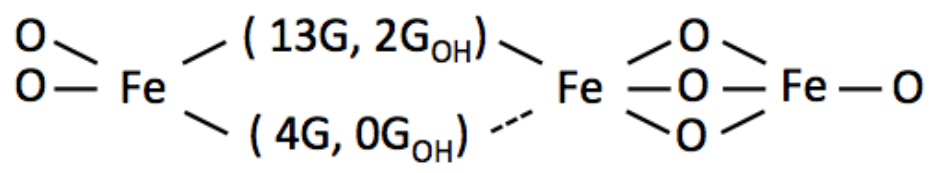 | 13 |
| 1349.537 | 35,711 | 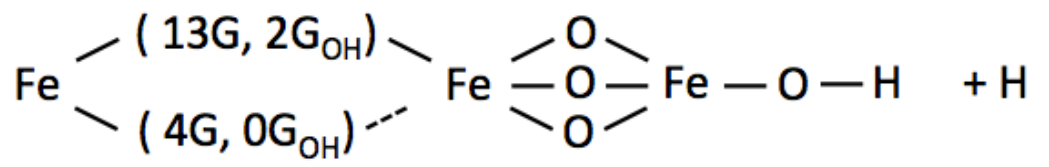 | 15 |
| 1105.326 | 21,809 | 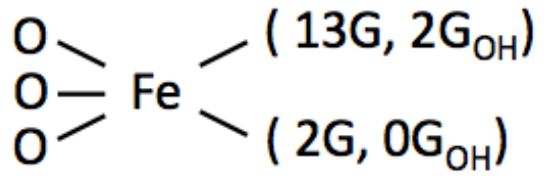 | 11 |
| 1080.392 | 47,283 | 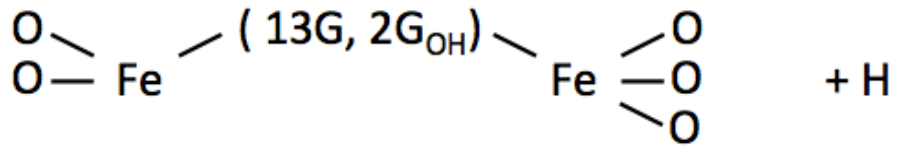 | 30 |
| 1065.435 | 21,632 | 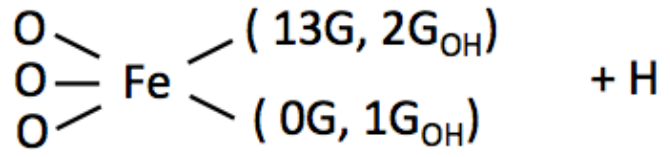 | 12 |
| 1051.598 | 12,989 | 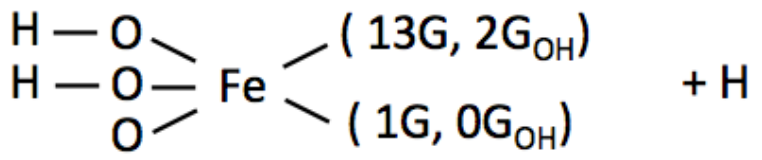 | 6 |
| 1025.364 | 33,140 | 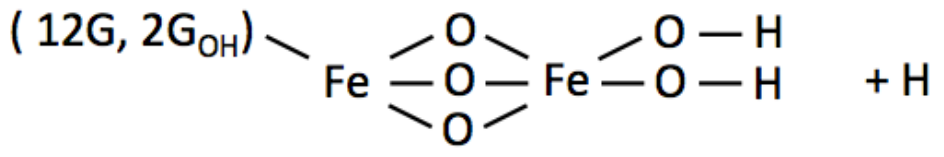 | 16 |
| 1009.431 | 15,864 | 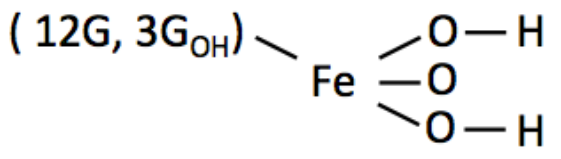 | 9 |
| 677.267 | 30,802 | $(8G, 3G_{OH}) + 2H$ | 23 |
| 675.283 | 10,140 | $(8G, 3G_{OH})$ | 8 |
| 661.289 | 9,488 | $(9G, 2G_{OH}) + 2H$ | 8 |
| 659.259 | 7,276 | $(9G, 2G_{OH})$ | 6 |
| 645.221 | 3,775 | $(10G, G_{OH}) + 2H$ | 3 |
| 643.292 | 5,982 | $(10G, G_{OH})$ | 5 |
| 547.197 | 4,694 | $(7G, 2G_{OH}) + 2H$ | 4 |
| 531.291 | 7,574 | $(8G, G_{OH}) + 2H$ | 7 |
| 515.476 | 4,117 | $(9G, 0G_{OH}) + 2H$ | 4 |
| 405.437 | 20,145 | 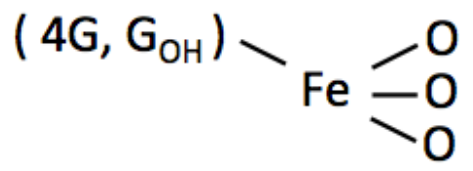 | 22 |
| 389.426 | 17,330 | 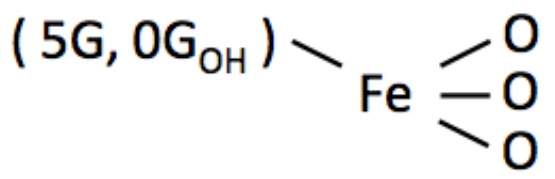 | 18 |
| 373.407 | 5,872 | 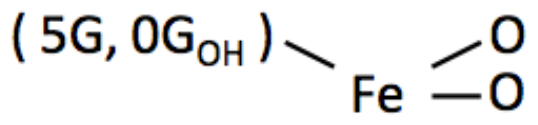 | 6 |
| 349.357 | 5,408 | 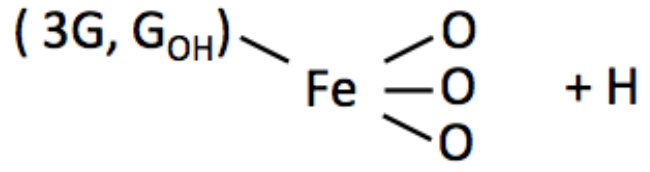 | 5 |
| 333.397 | 4,731 | 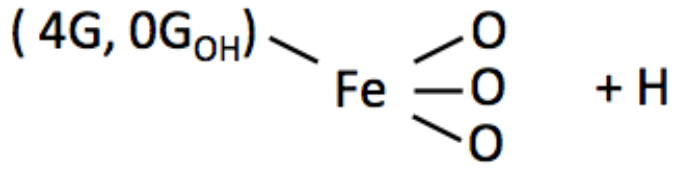 | 5 |

## S3.2 Additional chain length information

The main spectra with matrix CHCA carried an intense series of peaks in the 1000 < m/z < 1100 range, corresponding to peaks that have been reported previously [1] at a much poorer signal-to-noise ratio. A view of this range is presented in Figure S3.3.

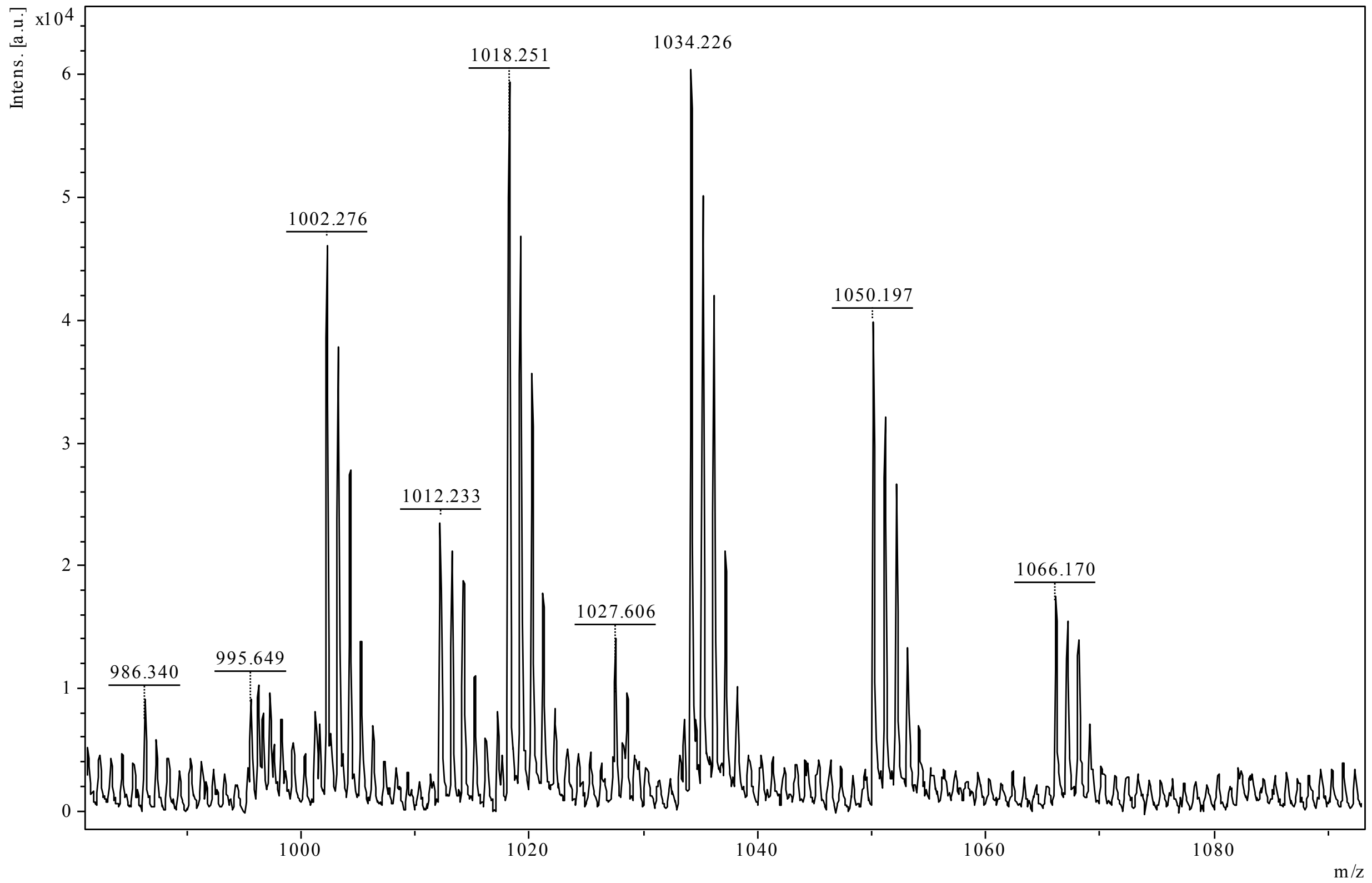

**Figure S3.3 Series of peaks ascending in oxygen count between m/z 1002 and 1066 from phase P2, matrix CHCA.**

These molecules do not contain Fe or Li, according to their lack of (-2) and (-1) satellite intensity, but they represent polymers of glycine and hydroxy-glycine as follows, in the notation (i,j,k) = (Gly, $Gly_{OH}$, Fe):
1066 = (11,6,0) + H
1050 = (12,5,0) + H
1034 = (13,4,0) + H
1018 = (14,3,0) + H
1002 = (15,2,0) + H
Together these peaks display a constant residue count of 17, suggesting that one well-defined fragment of hemolithin could be a peptide chain of that length.

An additional observation was made of a simple 17Gly chain at m/z 979, the largest peak in a spectrum of phase P2a (methods) that had been extracted for much longer than phase P2, viz 8 days at room temperature. This had the following characteristics in an isotope fit:

**Table S3.3 Isotope analysis of sample P2A CHCA at m/z 979.**

| **Obs. m/z Integer m/z** | **Satellite** | | | | | | | **S/N 0 peak** |
|---|---|---|---|---|---|---|---|---|
| 979.593 | (-3) | (-2) | (-1) | 0 | (1) | (2) | (3) | |
| Data | - | .008 | .038 | .497 | .307 | .115 | .026 | 61 |
| Fit | 0 | 0 | .039 | .497 | .311 | .114 | .031 | |
| **979** | (17,0,0) + Li + 3H 15,000/1,015 | | | | | | | |

The $^2$H enhancement of 15,000‰ was the lowest one in our observations, we believe because of a background rate of $^1$H / $^2$H substitution during this relatively long extraction in an aqueous system at room temperature.

In contrast, a 15-unit chain is suggested by the fragmentation of parent ion at m/z 2069 into the two most prominent peaks in phase P2, matrix SA, at 1027 and 1043. Each peak has intensity close to 200,000 and S/N = 111 and 114 respectively. The proposed fragmentation is illustrated in Figure S3.4
.

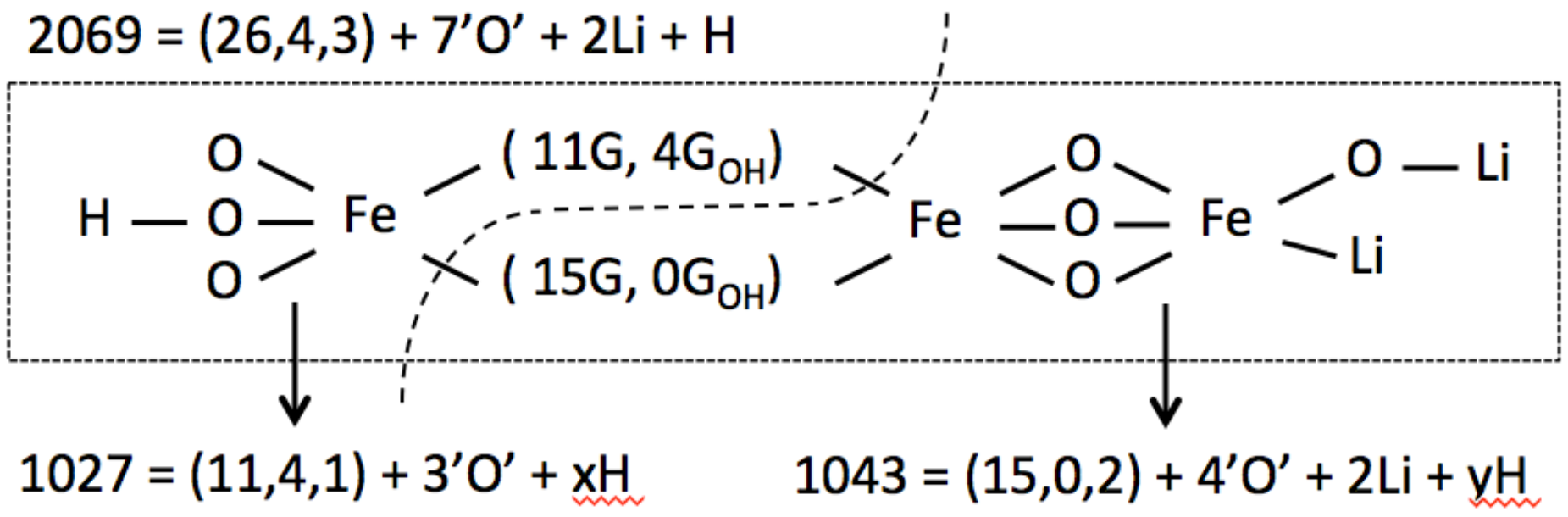

**Figure S3.4 Dominant fragments 1027 and 1043, with x + y = 2, indicate that a dominant break-up mode is via a) bonds of the peptide chains to Fe atoms, and b) unzipping of the hydrogen bonds between anti-parallel peptide chains.**

Lastly, a 16-unit chain is strongly suggested by the observation [2] of a monomer and dimers of 2320, described in S5.

In conclusion, there is evidence in the data for 15,16 and 17 residue glycine chains. In most cases several residues within any of these chains have been converted in a prior oxidizing environment to $Gly_{OH}$.

# S4 Map of Molecular Relationships

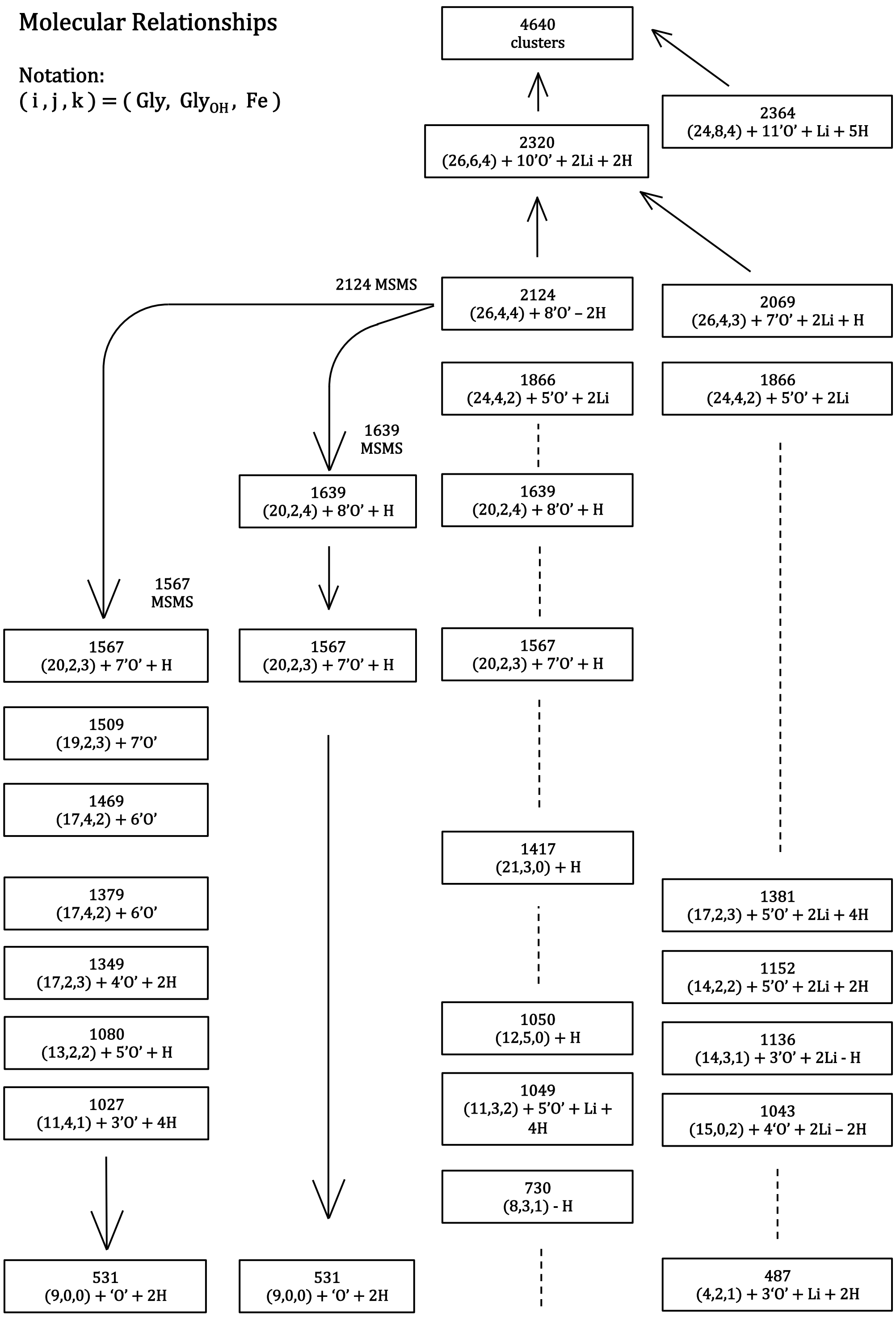

**Figure S4.1. Summary of molecular relationships. Left column, major peaks in the 1567 MS/MS spectrum. Left-center column: 1637 MS/MS, Right-center column: main CHCA spectrum peaks. Right column: Li-bearing molecule assignments.**

## S5. The 4641Da polymer peaks revisited.

The entities of chain length 16 Gly at 2320, 2364 and a 17Gly type at mass 2402 can explain the previously observed [2] high intensity complex of polymer amide peaks in the region of 4641Da (Figure S5.1). This complex was observed both in Acfer 086 and Allende. Not considering matrix adducts or clusters above 5,000Da we list in Table S5.1 the principal intense peaks from that data set [2], together with their new assignments.

**Table S5.1 Molecular assignments to the previously observed 4641Da complex.**

| Assignment | Observed m/z | Intensity | Calculated m/z |
|---|---|---|---|
| 2320 - Li | 2313.284 | 82,221 | 2313.377 |
| 2 x [2320 – (O + Li)] | 4595.116 | 49,287 | 4594.764 |
| 2 x [2320 – (O + Li)] + ‘O’ | 4611.582 | 49,804 | 4610.759 |
| 2 x [2320] + H | 4641.582 | 317,999 | 4641.794 |
| 2 x [2320] + ‘O’ | 4657.215 | 260,378 | 4656.780 |
| 2 x [2364 – Li + H] | 4716.357 | 48,394 | 4716.753 |
| 2 x [2364 + Li + H] | 4744.880 | 42,350 | 4744.818 |
| 2 x [2402] + H | 4805.988 | 42,823 | 4805.900 |
| 2 x [2402] + ‘O’ | 4821.152 | 49,821 | 4820.887 |

All the peaks are directly related to proposed 2320, 2364 and 2402 molecules. In the present series 2320 was not observed but its existence was derived from:
a) the (2320 – Li) observation in Table S5.1, and
b) the exact dimer mass of 2 x 2320 that was the highest peak in the 4641Da complex.
Higher mass clusters in multiples of 4641Da were also observed [2], suggesting that 2320 is being rapidly removed by dimer formation and complexing.

The proposed structures of 2124, 2320, 2364 and 2402 are shown below in Figure S5.2. The 2320 and 2364 molecules have in common a 16-residue per side anti-parallel chain but 2364 has an additional degree of oxidation to contain 4 hydroxy-glycines per side as opposed to 3 in 2320. Apart from this there is an additional ‘OH’ and one Li replaced by H. One of the 5H in 2364 is not drawn, as it is presumed to be a mass spectrometry adduct.

The 2402 molecule is based upon a 17Gly chain, with less oxidation than 2320 but otherwise similar end groups. It is present at 13% of the 2320 level.

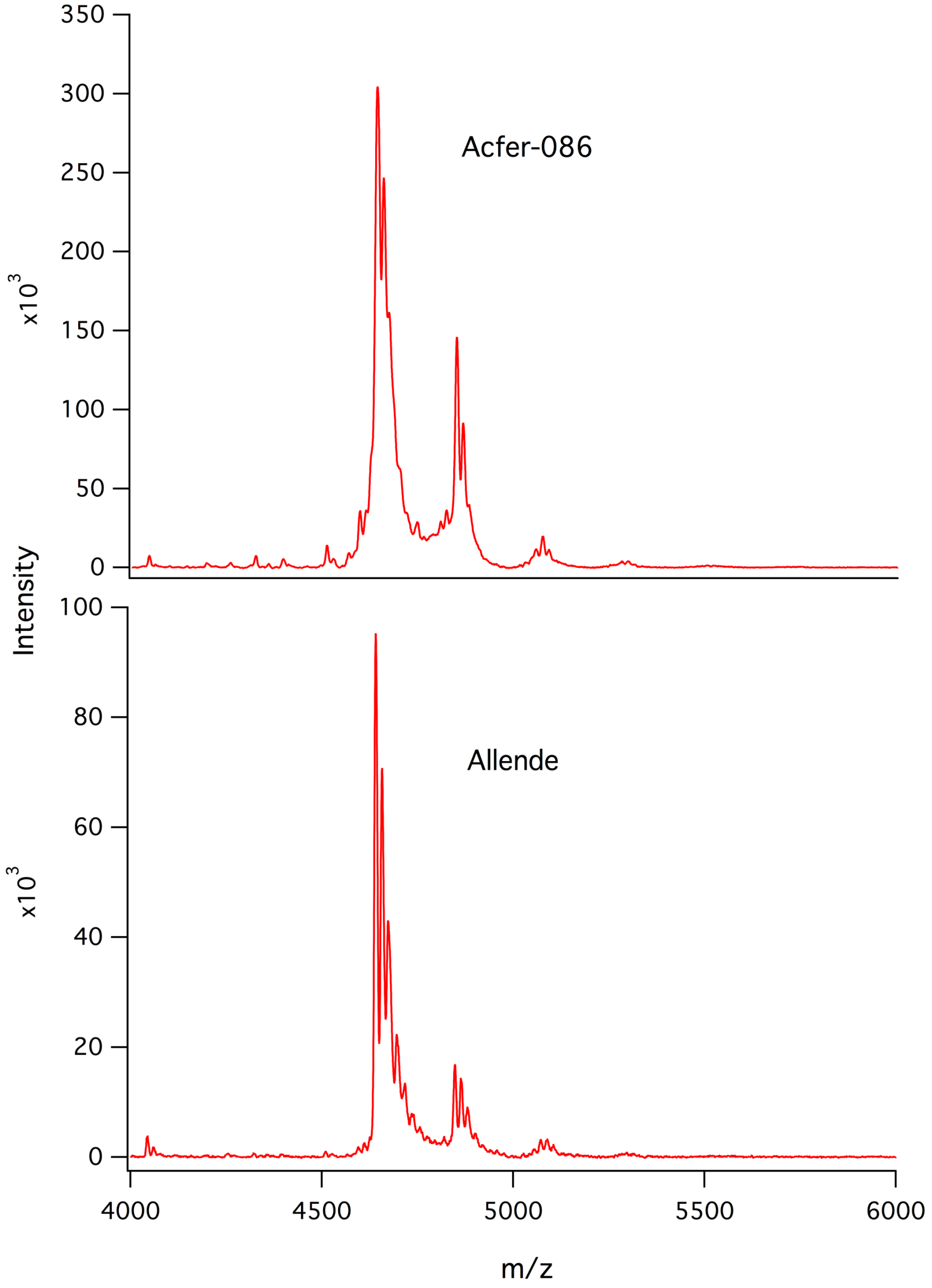

**Figure S5.1 Reproduced from [2]. Phase P2 mass spectra in the 4000 < m/z < 6000 range of Acfer 086 and Allende using sinapinic acid (SA) matrix.**

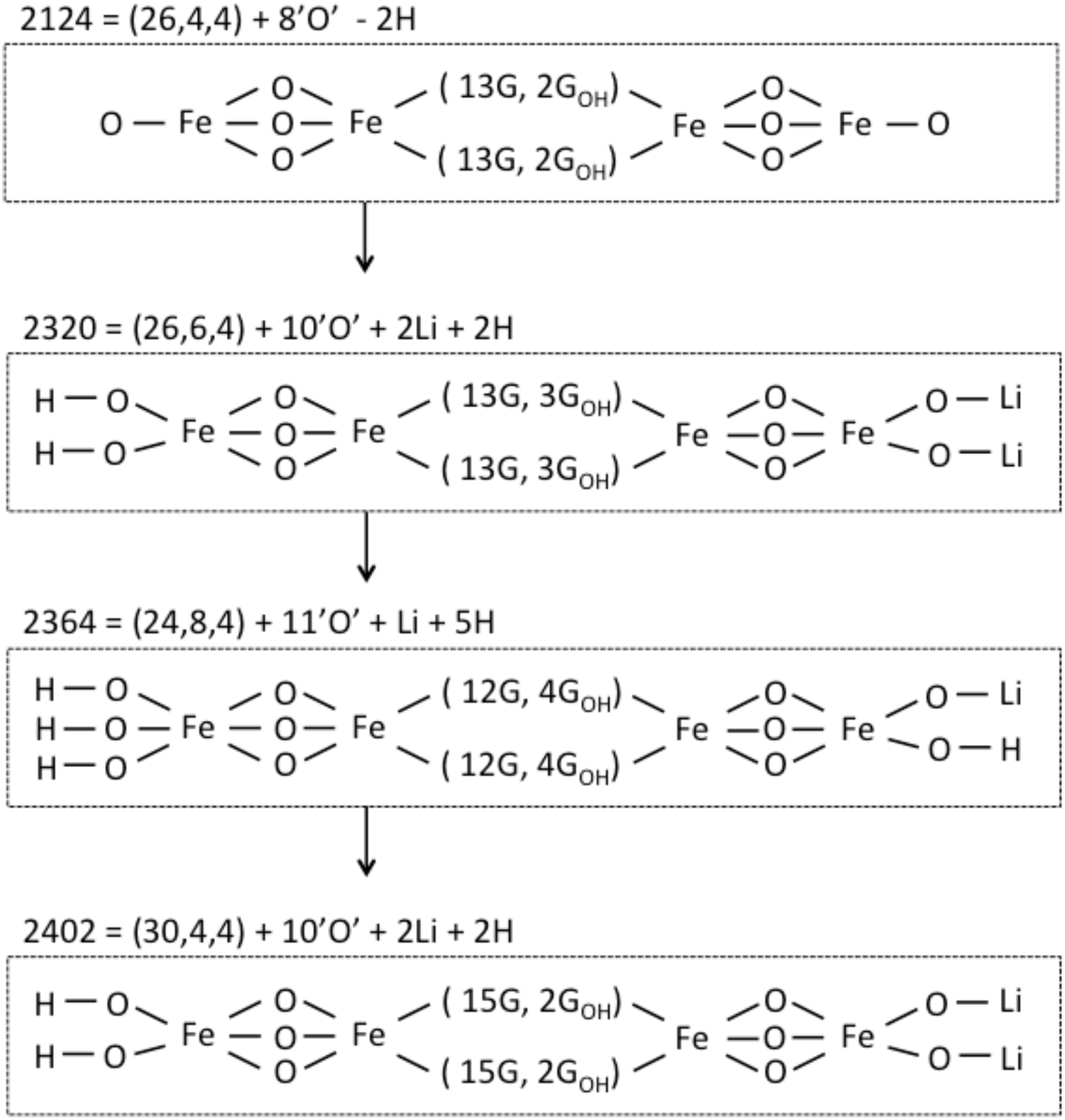

**Figure S5.2 Proposed 2124, 2320, 2364 and 2402 structures**